\documentclass[aps,pre,twocolumn,superscriptaddress,floatfix,notitlepage]{revtex4-2}
\usepackage[utf8]{inputenc}
\usepackage{amsmath,amsfonts,amssymb,amsthm,bm,color,dsfont,graphicx,psfrag,mathtools,cancel}
\usepackage{hyperref}
\usepackage{tikz}
\usetikzlibrary{positioning}
\usepackage{pgfplots}
\pgfplotsset{compat=1.17}
\usepackage{xcolor}
\usepackage{soul}

\usepackage[export]{adjustbox}
\usepackage{natbib}
\usepackage[hybrid]{markdown}
\usepackage{mathrsfs}

\usepackage{blkarray}
\usepackage{multirow}
\usepackage[caption = false]{subfig}

\begin{document}

\title{Integrable Model of a Superconductor with non-Fermi liquid and Mott Phases}

\author{Santhosh M}
\affiliation{Department of Physics, Indiana University, Bloomington, Indiana 47405, USA}

\author{Jorge Dukelsky}
\affiliation{Instituto de Estructura de la Materia, IEM-CSIC, Serrano 123, E-28006 Madrid, Spain}

\author{Gerardo Ortiz}
\affiliation{Department of Physics, Indiana University, Bloomington, Indiana 47405, USA}
\affiliation{Institute for Advanced Study, Princeton, NJ 08540, USA}

\date{\today}

\begin{abstract}
We present and analyze an exactly solvable interacting fermionic pairing model, which features interactions that entangle states at momenta $\mathbf{k}$ and $-\mathbf{k}$. These interactions give rise to novel correlated ground states, leading to a rich phase diagram that includes superconducting, multiple metallic, and Mott-insulating phases. At finite interaction strengths, we observe the emergence of multiple many-body Fermi surfaces, which violate Luttinger's theorem and challenge the conventional Landau-Fermi liquid paradigm. A distinguishing feature of our model is that it remains quantum integrable, even with the addition of pairing interactions of various symmetries, setting it apart from the Hatsugai-Kohmoto model. Our results provide an analytically tractable framework for studying strong correlation effects that give rise to fractionalized excitations and unconventional superconductivity, offering valuable insights into a broad class of integrable many-body systems.
\end{abstract}
\maketitle

\section{Introduction}
\label{Section1}

Strongly correlated electron systems exhibit a rich spectrum of exotic quantum phases, including Mott insulators, unconventional metals and superconductors \cite{Imada1998, Kotliar2004, Sachdev1993, Sachdev2010,Chowdhury2022, Varma1989, Coleman2005, Rice2012, Zaanen2019, Li2022}. Yet, their theoretical understanding remains a major challenge, largely due to the absence of exact solutions, necessitating reliance on numerical methods and approximate effective theories \cite{Lee2006, Georges1996}. In this context, quantum integrable models serve as indispensable tools for probing the structure of quantum matter and for developing intuition about non-perturbative phenomena. These models often uncover hidden features— such as infinite families of conserved charges, dualities, and exact symmetries— that shed light on the organizing principles of complex many-body systems.

One of the field’s central theoretical challenges is to realize a number-conserving fermionic system that encompasses non-Fermi liquid behavior, Mott insulating phases, and superconductivity within a coherent framework. This challenge is heightened by the seemingly antagonistic nature of the mechanisms underlying these competing phases: superconductivity typically arises from effective electron pairing, driven by attractive interactions, whereas non-Fermi liquids and Mott insulators emerge from strong electron repulsion and the potential breakdown of Landau quasiparticle descriptions. In this paper, we construct an exactly solvable model of interacting fermions that realizes all three paradigmatic competing phases— Mott-insulating, non-Fermi liquid, and superconducting— within a single, unified framework. Our model offers a rare analytic platform to explore the competition and coexistence of these distinct phases.

Our construct belongs to the class of Richardson-Gaudin models ~\cite{Richardson1963,Gaudin1976,Class2001,Ortiz2005,claeys_richardson-gaudin_2018}, with its normal (non-pairing) phase inspired by the Hatsugai-Kohmoto (HK) model \cite{Hatsugai1992,Hatsugai1996, Zhao2022, Zhao2023, Mai2023, Tenkila2025, Zhao_2025}, which is notable for violating Luttinger’s theorem \cite{Oshikawa2000, Luttinger1960} and realizing emergent Mott physics. In the absence of pairing interactions, our model and the HK model share a common eigenbasis, reflecting a deep structural connection. However, our model exhibits a significantly richer phase diagram, featuring ten distinct metallic phases and three inequivalent Mott-insulating phases, each distinguished by the filling of different fermionic multiplets. These key differences allow the inclusion of number-conserving superfluid pairing interactions while preserving exact integrability. However, it introduces a key trade-off: while the HK model is separable at each point in the Brillouin zone (BZ), our model is only separable in momentum pairs $(\mathbf{k}, -\mathbf{k})$. This subtle distinction gives rise to a nontrivial entanglement structure in the eigenstates— even in the absence of pairing— reflecting an intrinsic momentum-space coupling encoded by the pseudospin representation.

As discussed in Section \ref{Section2}, our model describes spinful fermions with competing attractive and repulsive interactions and exhibits a rich symmetry structure, including translational invariance, fermion number conservation, a local $su(2)$ \textit{gauge}, and particle-hole symmetries. We establish the model's quantum integrability and determine its full many-body spectrum with algebraic complexity, revealing a macroscopic ground-state degeneracy. Interestingly, the non-interacting Fermi liquid at half-filling possesses an additional $\mathbb{Z}_2$ spin-resolved particle-hole symmetry \cite{ AndersonHaldane2001,Zhao2022, Phillips2020}, whose operator is amenable to exact description. This symmetry is explicitly broken in the interacting case, and its breaking serves as a diagnostic for non-Fermi liquid behavior in our model.

One of the most striking features of our model, shared with the HK model, is the emergence of multiple {\it many-body Fermi surfaces} \cite{Senthil2008}, even at arbitrarily small but finite repulsive interaction strengths. As detailed in Section~\ref{Section3}, this structure is uncovered through an exact analysis of the quantum phase diagram, carried out in the absence of pairing and in an arbitrary number of spatial dimensions $d$. We determine the precise locations of the quantum phase transitions and find that each non-analyticity in the ground-state energy corresponds to a many-body Fermi surface intersecting the boundary of the BZ. In this specific sense, the interaction-driven quantum phase transitions in our model can be interpreted as many-body analogues of Lifshitz transitions~\cite{Lifshitz1960,Volovik2017}, wherein the topology of the Fermi surface is reshaped by electronic correlations rather than by band structure alone. Remarkably, this mechanism gives rise to a cascade of unconventional metal–to-Mott insulator transitions, whose behavior we are able to characterize analytically. Perhaps unsurprisingly, a mathematical connection emerges between the HK model and our own. Specifically, the ground-state energy of our Hamiltonian, when projected onto the subspace annihilated by the momentum pseudospin operators, coincides with that of the HK model for rescaled  values of the repulsive interaction strength and fermion density. This correspondence holds irrespective of both the interaction strength and the spatial dimensionality of the system.

We now turn to a fundamental question: {\it What is the nature of the normal state excitations in our model?} This question is addressed in part in Section~\ref{Section4}, where we compute the exact single-particle retarded Green's function and the density of states. Strikingly, the Green's function exhibits four distinct poles, each located at the position of one of the four many-body Fermi surfaces. This feature indicates that the quasiparticles in our model possess infinite lifetimes and behave as unconventional excitations carrying {\it fractional} quantum numbers— a characteristic also present in the HK model, albeit with only two distinct poles corresponding to holons and doublons. Despite the intricate nature of the quasiparticle excitations, the structure of the corresponding quasi-pairs is remarkably simple: they are given exactly by the pseudospin raising and lowering operators. 

A detailed exploration of how excitation fractionalization influences the system’s thermodynamic behavior is presented in Section~\ref{Section5}. For instance, at zero temperature, the charge compressibility decomposes into a linear combination of four free-fermion contributions, each resembling a Hubbard-like band centered at a distinct value of the repulsive interaction. In the strongly repulsive regime and across arbitrary spatial dimensions, we demonstrate the existence of four distinct zero-temperature non-Fermi liquid phases. Each phase features a unique many-body Fermi surface whose location is determined solely by the fermion density, remaining entirely independent of the interaction strength—mirroring the behavior of a non-interacting Fermi liquid. Strikingly, an {\it emergent particle-hole symmetry} appears in each phase as a direct consequence of fractionalization, highlighting an unexpected and fundamental property of the system. In the non-Fermi liquid phases, the specific heat exhibits linear temperature scaling at low temperatures, while in the Mott insulating phases it is exponentially suppressed, consistent with a nonzero charge gap. Finally, the macroscopic ground-state degeneracy gives rise to a finite residual entropy at low temperatures, regardless of spatial dimensionality or interaction strength.

A comprehensive understanding of the normal state necessitates a thorough investigation of its potential superconducting instability. In Section~\ref{Section6}, we examine the formation of a bound electron pair in the presence of an arbitrarily weak attractive interaction. To probe the instability of the various non-Fermi liquid and Mott insulating phases, we compute the binding energy of a single Cooper pair \cite{Cooper-pair}, taking into account the repulsive interaction. For the one-dimensional case, we derive closed-form expressions that reveal distinct scaling behaviors depending on the nature of the underlying normal state. Some phases exhibit essential singularities in the weak-attraction limit, signaling a strong pairing tendency, while others display regular behavior, possibly indicating the absence of a superconducting instability. A full analysis of the resulting superconducting phases is deferred to future work. 

We conclude with a series of appendices that present the detailed calculations underlying the main results. While the core of the paper focuses on $s$-wave singlet pairing, the model admits extensions to other pairing symmetries while retaining quantum integrability. In Appendix~\ref{AppendixE}, we demonstrate one such extension by incorporating $p_x + i p_y$ superconducting terms. A comparison with the quantum phase diagram of the HK model is presented in Appendix~\ref{AppendixHK}, highlighting key similarities and differences. The analytic calculation of the retarded Green's function, presented in Appendix~\ref{AppendixB}, is particularly elegant and illuminating. Finally, Appendix~\ref{Onepairapp} provides the details of the Cooper-pair instability analysis, which are essential for computing the binding energy.



\section{Model Hamiltonian}
\label{Section2}

Our interacting Hamiltonian can be written in terms of fermion creation (annihilation) operators $c^\dagger_{\bf{k}\sigma}$ ($c^{\;}_{\bf{k}\sigma}$), of momentum ${\bf{k}}=(k_1,k_2,\cdots, k_d)$ and spin $\sigma=\uparrow,\downarrow$, defined on a $d$-dimensional lattice (with lattice constant $a=1$). It consists of a {\it normal state} part
\begin{equation}
H_{\sf n} = \sum_{\bf{k}} \tilde \epsilon_{\bf{k}} \,\hat N_{\bf{k}} +  H_{\sf U},
\label{nonfermi}
\end{equation}
with
\begin{equation}
H_{\sf U} =   U \sum_{\bf{k}} \left ( \tau_{\bf{k}}^+ \tau_{\bf{k}}^- + \frac{\hat N_{\bf k}(\hat N_{\bf k}-1)}{4} \right ),
\label{nonfermi}
\end{equation}
where $\hat N_{\bf k}=n_{\bf{k} \uparrow}+ n_{\bf{k} \downarrow} +n_{-\bf{k} \uparrow}+ n_{-\bf{k} \downarrow}$, with $n_{\bf{k}\sigma}=c^\dagger_{\bf{k}\sigma}c^{\;}_{\bf{k}\sigma}$, $\tilde \epsilon_{\bf{k}}=\epsilon_{\bf{k}}-\mu$, with $\epsilon_{\bf{k}}$ representing the band dispersion and $\mu$ the chemical potential, $U$ the interaction strength, and {\it throughout the paper the summations and products extend over all $\bf k$ vectors with $k_1 > 0$}. (Without loss of generality we assume that the number of lattice sites along an arbitrary spatial direction is even, which we can take to be $2L_i$, $i=1,\cdots,d$, with total volume $V=\prod_{i=1}^d (2L_i)$). The resulting set of allowed momenta along each spatial direction is ${\cal S}_{k_i}=(2L_i)^{-1}\{\pm \pi,\pm 3\pi, \cdots, \pm \pi(2L_i-1) \}$. 
The operators $\tau^+_{\bf{k}} = c_{{\bf{k}} \uparrow}^{\dagger} c_{-\bf{k} \downarrow}^{\dagger} - c_{\bf{k} \downarrow}^{\dagger} c_{-\bf{k} \uparrow}^{\dagger} = (\tau^-_{\bf{k}})^{\dagger}$ and $\tau^z_{\bf{k}} = \frac{\hat N_{\bf k}}{2}-1$ generate a pseudospin $su(2)$ algebra
\begin{equation}
\left[ \tau_{\bf{k}}^+, \tau_{\bf{k}}^- \right] = 2 \tau_{\bf{k}}^z ; \quad \left[ \tau_{\bf{k}}^z, \tau_{\bf{k}}^{\pm} \right] = \pm \tau_{\bf{k}}^{\pm} ,
\end{equation}
with Casimir operator ${\cal C}^2_{\bf k}=\tau_{\bf{k}}^+ \tau_{\bf{k}}^- +\tau_{\bf{k}}^z (\tau_{\bf{k}}^z-1)$.
In addition, our two-body model Hamiltonian
\begin{eqnarray}
H= H_{\sf n} + H_{\sf p}  ,
\label{Hamiltoniansw}
\end{eqnarray}
contains a pairing interaction of strength $G$. To aid understanding, this section focuses on s-wave superconducting terms of the form.\begin{equation}
H_{\sf p} = -G\sum_{\bf{k}, \bf{k'}}\tau_{\bf{k}}^+\tau_{\bf{k'}}^- \ ,
\end{equation}

\subsection{Symmetries of the Hamiltonian}
\label{Symmetries of the Hamiltonian}

\subsubsection{Fermion number conservation}

It is straightforward to show that the total 
fermion number operator
\begin{equation}
\hat{N} = \sum_{\bf{k}} \hat N_{\bf k},
\label{eq:N_operator}
\end{equation}
commutes with the Hamiltonian, i.e., $[ H,\hat{N}]=0$

\subsubsection{Translational Symmetry}

The operator 
\begin{eqnarray}
\hat{T}=\sum_{\bf k} {\bf k} \, (n_{\bf{k} \uparrow}+ n_{\bf{k} \downarrow} -n_{-\bf{k} \uparrow}- n_{-\bf{k} \downarrow}) = \sum_{\bf k} \hat{T}_{\bf k} ,
\end{eqnarray}
represents translations and commutes with $H$.

\subsubsection{Spin $su(2)$ Gauge Symmetry}

The spin $su(2)$ algebra with generators $S^+_{\bf{k}} = c_{{\bf{k}} \uparrow}^{\dagger} c_{-\bf{k} \downarrow}^{\;} + c_{-\bf{k} \uparrow}^{\dagger} c_{\bf{k} \downarrow}^{\;} = (S^-_{\bf{k}})^{\dagger}$ and $S^z_{\bf{k}} = \frac{1}{2}(n_{\bf{k} \uparrow}+n_{-\bf{k} \uparrow}-n_{\bf{k} \downarrow} - n_{-\bf{k} \downarrow})$, 
\begin{equation}
\left[ S_{\bf{k}}^+, S_{\bf{k}}^- \right] = 2 S_{\bf{k}}^z ; \quad \left[ S_{\bf{k}}^z, S_{\bf{k}}^{\pm} \right] = \pm S_{\bf{k}}^{\pm} ,
\end{equation}
commutes with the pseudospin $su(2)$ algebra, meaning that it constitutes a gauge symmetry of our model Hamiltonian $H$. 

\subsubsection{Particle-Hole Symmetry}

To investigate the effect of particle-hole-exchange in our model, we explicitly express the fermion operators in position ${\bf x}=(x_1, x_2,\cdots, x_d)$ (real-)space as
\begin{eqnarray}
c_{{\bf{x}} \sigma} &=& \frac{1}{\sqrt{V}} \sum_{\bf{k}} ( \mathrm{e}^{-i \bf{k} \cdot \bf{x}}+\mathrm{e}^{i \bf{k} \cdot \bf{x}} )c_{\bf{k} \sigma},
\end{eqnarray}
where $x_i = 1, \cdots, 2L_i$.  Our system is defined on a {\it bipartite lattice}, composed of two interpenetrating lattices $\mathrm{A}$ and $\mathrm{B}$,  with an even number of lattice sites in each spatial dimension. Specifically, we shall adopt the convention $\bf{x}\in \mathrm{A}$ if $\sum_{i=1}^d x_i$ is odd and $\bf{x} \in \mathrm{B}$, otherwise. The unitary operator implementing the particle-hole transformation of interest is given by $\mathcal{K}_{ph}= \mathcal{K_{\uparrow}}\mathcal{K_{\downarrow}}$. The \textit{spin-resolved particle-hole transformation} $\mathcal{K_{\sigma}}$ acts selectively on the spin sector \( \sigma \), while leaving the opposite spin \( \bar{\sigma} \) unchanged. In terms of fermion operators it can be expressed as \cite{BatistaOrtiz2004}
\begin{equation}
\mathcal{K_{\sigma}}=\prod_{\bf{x}} \left( \exp \left[i \pi \delta_{\bf{x} \mathrm{B}} \hat{n}_{\bf{x} \sigma}\right] \exp \left[i \frac{\pi}{2}\left(c_{\bf{x} \sigma}^{\dagger}+c_{\bf{x} \sigma}\right)\right] \right),
\label{particlehole}
\end{equation}
where $\delta_{\bf{x} \mathrm{B}}$ is one if $\bf{x} \in \mathrm{B}$ and zero, otherwise. Notice that the above expression only defines $\mathcal{K}_{\sigma}$ up to a sign depending upon the ordering of the product. This ambiguity does not affect the results below, as long as the ordering is consistent throughout the calculation. It acts as
\begin{equation}
\mathcal{K}_{\sigma} c_{\bf{x}\sigma}^{\dagger} \mathcal{K}_{\sigma}^{\dagger}=\left\{\begin{array}{cc}
-c_{\bf{x} \sigma} & \text {if $ {\bf{x}}\in \mathrm{A} $  }  \\
c_{\bf{x}\sigma} & \text {if $ {\bf{x}}\in \mathrm{B} $  } 
\end{array} ,\right.
\end{equation}
and $\mathcal{K}_{\sigma} c_{\bf{x}\bar \sigma}^{\dagger} \mathcal{K}_{\sigma}^{\dagger}=c_{\bf{x}\bar \sigma}^{\dagger}$, because the number of lattice sites is always even. 

Consequently, one can prove that $\mathcal{K}_{\sigma}$ acts upon the momentum-space fermion operators as 
\begin{eqnarray}
\mathcal{K}_{\sigma}c_{\bf{k} \sigma}\mathcal{K}_{\sigma}^{\dagger} &=&  c_{\bf{-k+K}, \sigma}^{\dagger} , \nonumber\\
\mathcal{K}_{\sigma}c_{\bf{k} \sigma}^{\dagger}\mathcal{K}_{\sigma}^{\dagger} &=&  c_{\bf{-k+K}, \sigma} ,
\end{eqnarray}
where $ {{K}}_i =  \text{sgn}(k_i) \pi $ for \( i = 1,2, \dots, d \).

If the dispersion relation satisfies $\epsilon_{\bf{k}} = - \epsilon_{\bf{-k +K}}$, $\mathcal{K}_{ph}$ acts on the Hamiltonian as 
\begin{eqnarray} 
\mathcal{K}_{ph}H_{\sf n}\mathcal{K}_{ph}^{\dagger} &=& H_{\sf n} +\left(2\mu-\frac{5}{2}U\right) \sum_{\bf{k}}(\hat{N}_{\bf{k}}-2) , \nonumber \\
\mathcal{K}_{ph}H_{\sf p}\mathcal{K}_{ph}^{\dagger} &=& H_{\sf p} + G\sum_{\bf{k}}(\hat{N}_{\bf{k}}-2) .
\end{eqnarray}

Notice that the equations above can be rewritten as  
\begin{equation}
\mathcal{K}_{ph}H|_{\mu}\mathcal{K}_{ph}^{\dagger} = H|_{-\mu + \frac{5U}{2}-G}  -aV
\label{parthole}
\end{equation}
where $2a= 4\mu -5U+2G$ is a constant. The above transformation ensures that the effect of the particle-hole operator $\mathcal{K}_{ph}$ on the Hamiltonian is to change the value of the chemical potential, up to a constant. Consequently, our model is particle-hole \textit{symmetric} about the half-filling point defined by $4\mu = 5U -2 G$.

\subsection{Classification of basis states}
\label{Classification of basis states}

Basis states of the Fock space \({\cal F} = \bigoplus_{N=0}^{\infty} \hat{\cal A} {\cal H}^{\otimes N} = \mathbb{C} \oplus {\cal H} \oplus \hat{\cal A} ({\cal H} \otimes {\cal H}) \oplus \cdots\), where \(\hat{\cal A}\) is the antisymmetrizer and \({\cal H}\) is the single-particle Hilbert space, can be classified by focusing on the 16-dimensional subspace ${\cal H}_{\bf k}$ spanned by states with up to four particles in fixed \(\mathbf{k}\) and \(-\mathbf{k}\) modes. The 16 states are specified in Table \ref{table1}.
\begin{table}[htb]
\[
\begin{array}{|c|c|c|c|c|}
\hline
 & M_{\bf k}, s^z_{\bf k}, T_{\bf k}, \nu_{\bf k} & \text{State}& \text{$\tilde E_{\bf k}(\mu,U)$}\\
\hline
1 & 0, 0, 0, 0 & |0\rangle & 0 \\
2 & 0, \frac{1}{2}, {\bf k}, 1 & c_{{\bf k}\uparrow}^\dagger  |0\rangle & \tilde\epsilon_{\bf{k}}\\
3 & 0, -\frac{1}{2}, {\bf k}, 1 & c_{\bf k  \downarrow}^\dagger |0\rangle &\tilde\epsilon_{\bf{k}}\\
4 & 0, \frac{1}{2}, -{\bf k}, 1 & c_{-{\bf k}\uparrow}^\dagger  |0\rangle & \tilde\epsilon_{\bf{k}}\\
5 & 0, -\frac{1}{2}, -{\bf k}, 1 & c_{-{\bf k}\downarrow}^\dagger  |0\rangle & \tilde\epsilon_{\bf{k}} \\
6 & 0, 0, 2{\bf k}, 2 & c_{\bf k \uparrow}^\dagger  c_{\bf k  \downarrow}^\dagger |0\rangle & 2\tilde\epsilon_{\bf{k}} + \frac{U}{2}\\
7 & 0, 1, 0, 2 & c_{\bf k \uparrow }^\dagger c_{-{\bf k}\uparrow}^\dagger  |0\rangle & 2\tilde\epsilon_{\bf{k}}+\frac{U}{2}\\
8 & 0, -1, 0, 2 & c_{\bf k  \downarrow}^\dagger c_{-{\bf k}\downarrow}^\dagger  |0\rangle &2\tilde\epsilon_{\bf{k}}+\frac{U}{2} \\
9 & 0, 0, -2{\bf k}, 2 & c_{-{\bf k}\downarrow}^\dagger  c_{-{\bf k}\uparrow}^\dagger  |0\rangle & 2\tilde\epsilon_{\bf{k}}+\frac{U}{2}\\
10 & 0, 0, 0, 2 & \frac{1}{\sqrt{2}}(c_{\bf k \uparrow}^\dagger  c_{-{\bf k}\downarrow}^\dagger  + c_{\bf k \downarrow}^\dagger  c_{-{\bf k} \uparrow}^\dagger ) |0\rangle & 2\tilde\epsilon_{\bf{k}}+\frac{U}{2}\\
11 & 1, 0, 0, 0 & \frac{1}{\sqrt{2}}\tau_{\bf k}^+ |0\rangle & 2\tilde\epsilon_{\bf{k}}+\frac{5U}{2}\\
12 & 1, \frac{1}{2}, {\bf k}, 1 & \tau_{\bf k}^+ c_{\bf k \uparrow}^\dagger  |0\rangle & 3\tilde\epsilon_{\bf{k}}+\frac{5U}{2}\\
13 & 1, -\frac{1}{2}, {\bf k}, 1 & \tau_{\bf k}^+ c_{\bf k \downarrow}^\dagger  |0\rangle & 3\tilde\epsilon_{\bf{k}}+\frac{5U}{2}\\
14 & 1, \frac{1}{2}, -{\bf k}, 1 & \tau_{\bf k}^+ c_{-{\bf k}\uparrow}^\dagger  |0\rangle & 3\tilde\epsilon_{\bf{k}}+\frac{5U}{2}\\
15 & 1, -\frac{1}{2}, -{\bf k}, 1 & \tau_{\bf k}^+ c_{-{\bf k}\downarrow }^\dagger |0\rangle & 3\tilde\epsilon_{\bf{k}}+\frac{5U}{2}\\
16 & 2, 0, 0, 0 & \frac{1}{2}\tau_{\bf k}^+ \tau_{\bf k}^+ |0\rangle & 4\tilde\epsilon_{\bf{k}}+5U\\ 
\hline
\end{array}
\]
\caption{Table of (normalized) basis states and associated $H_{\sf n}$ energies. Here, $2M_{\bf k}+\nu_{\bf k}=N_{\bf k}$.}
\label{table1}
\end{table}

In this way, a complete set of many-body (un-normalized) basis states can now be written as
\begin{eqnarray}
|\{M_{\bf k}, s^z_{\bf k}, T_{\bf k}, \nu_{\bf k}\}\rangle = \prod_{\bf k} \left( \tau_{\bf k}^+ \right)^{M_{\bf k}} | \lambda_{\bf k} \rangle ,
\end{eqnarray}
where $| \lambda_{\bf k} \rangle$ denotes a state in Table \ref{table1} characterized by $M_{\bf k}=0$ (i.e., $\tau_{\bf k}^- | \lambda_{\bf k} \rangle = 0$). These states can be distinguished from each other by the quantum numbers: $S^z_{\bf k} |\lambda_{\bf k} \rangle=s^z_{\bf k}|\lambda_{\bf k} \rangle$, $\hat T_{\bf k} |\lambda_{\bf k} \rangle=T_{\bf k}|\lambda_{\bf k} \rangle$, $\hat N_{\bf k} |\lambda_{\bf k} \rangle=\nu_{\bf k}|\lambda_{\bf k} \rangle$. The energies $\tilde E_{\bf k}(\mu,U)$ can be compactly written as
\begin{eqnarray} \hspace*{-0.5cm}
\tilde E_{\bf k}(\mu,U)=(2 \tilde \epsilon_{\bf k}+\frac{5}{2}U) M_{\bf k}+\tilde \epsilon_{\bf k} \nu_{\bf k}+\frac{U}{4}\nu_{\bf k}(\nu_{\bf k}-1) .
\label{spenergies}
\end{eqnarray}
Note that quantum numbers $s^z_{\bf k}$ and $T_{\bf k}$ do not appear in this expression, meaning that there will be macroscopic degeneracies associated to those symmetries (see Sec. \ref{Macroscopic Degeneracy}).

In the following subsections, we examine the formal aspects of this model in detail.

\subsection{Quantum Integrability}
\label{Quantum Integrability}

We next show that the $s$-wave Hamiltonian \eqref{Hamiltoniansw} is exactly solvable, i.e., the full spectrum can be determined with algebraic complexity. First, we note that our model Hamiltonian  $H$, when $U=0$, reduces to the pairing Hamiltonian solved by Richardson in the sixties \cite{Richardson1963,Richardson64}
\begin{equation}
H\left(  U=0\right)  =\sum_{\mathbf{k}}\tilde{\epsilon}_{\mathbf{k}
}\hat{N}_{\mathbf{k}}-G\sum_{\mathbf{k,k}^{\prime}}\tau_{\mathbf{k}}
^{+}\tau_{\mathbf{k}^{\prime}}^{-}\text{ .}
\label{H(U=0)}
\end{equation}
A product state of pair wavefunctions was proposed as an ansatz for the eigenvectors
\begin{equation}
\left\vert \Psi_{M,\lambda}\right\rangle =
{\displaystyle\prod\limits_{\alpha=1}^{M}}
B_{\alpha}^{\dagger}\left\vert \lambda\right\rangle \text{, \ }B_{\alpha
}^{\dagger}=\sum_{\mathbf{k}}\frac{1}{\tilde{\epsilon}_{\mathbf{k}%
}-x_{\alpha}}\tau_{\mathbf{k}}^{+},
\label{Ansatz}
\end{equation}
where $x_{\alpha}$, $\alpha =1, \cdots, M$,  are the pairon energies and $\left\vert \lambda\right\rangle=\bigotimes_{\bf{k}} | \lambda_{\bf{k}} \rangle$ are the \textit{pairon vacua}, which are annihilated by all the $\tau_{\bf k}^-$.

In order to fulfill the eigenvalue equation%
\begin{eqnarray}
H\left(  U=0\right)  \left\vert \Psi_{M,\lambda}\right\rangle =E\left(
\mu,G,M,\{ \nu_{\bf k} \}\right)  \left\vert \Psi_{M,\lambda}\right\rangle ,
\end{eqnarray}
for a given pairon vacuum $\left\vert \lambda \right\rangle$ with $\left\{  \nu_{\mathbf{k}}\right\}  $
unpaired fermions for each momentum pair $(\mathbf{k}, -\mathbf{k})$, the $M$ pairon energies
$x_{\alpha}$ must satisfy the set of $M$ Richardson equations
\begin{equation}
1+G\sum_{\mathbf{k}}\frac{\frac{\nu_{\mathbf{k}}}{2}-1}{\tilde{\epsilon
}_{\mathbf{k}}-x_{\alpha}}-G\sum_{\alpha\neq\beta}\frac{1}{x_{\alpha}
-x_{\beta}}=0\text{ .} \label{Equations}%
\end{equation}
The corresponding eigenvalues are
\begin{equation}
E\left(\mu, G, M, \{\nu_{\bf k}\}\right)  =\sum_{\mathbf{k}}\tilde{\epsilon}_{\mathbf{k}}
\nu_{\mathbf{k}}+2\sum_{\alpha}x_{\alpha}\text{ .}
\end{equation}

We now consider the inclusion of the repulsion term, $H_{\sf U}$.  It turns out that $H_{\sf U}$ commutes with $H\left(  U=0\right)  $, and therefore, it is a constant of motion. To determine its spectral contribution consider the action on the ansatz state
\begin{eqnarray}
H_{\sf U}\left\vert \Psi_{M,\lambda}\right\rangle &=&\sum_{\alpha=1}^{M}
{\displaystyle\prod\limits_{\beta\left(  \neq\alpha\right)  }}
B_{\beta}^{\dagger}\left[  H_{\sf U},B_{\alpha}^{\dagger}\right]  \left\vert
\lambda\right\rangle +
{\displaystyle\prod\limits_{\alpha=1}^{M}}
B_{\alpha}^{\dagger}H_{\sf U}\left\vert \lambda\right\rangle \nonumber \\
&=& \left ( \frac{5U}{2}M+\frac{U}{4}\sum_{\mathbf{k}}
\nu_{\mathbf{k}}\left(  \nu_{\mathbf{k}}-1\right)\right )\! \left\vert \Psi_{M,\lambda}\right\rangle   ,
\end{eqnarray}
where $H_{\sf U}|\lambda\rangle =\frac{1}{4}U\sum_{\mathbf{k}}\nu_{\mathbf{k}}\left(  \nu_{\mathbf{k}}-1\right)$ and $\left[  H_{\sf U},B_{\alpha}^{\dagger}\right]  \!=\frac{5}{2}UB_{\alpha}^{\dagger}$.

Consequently, the complete set of eigenstates of the Hamiltonian \eqref{Hamiltoniansw} is given by the Richardson ansatz \eqref{Ansatz}, which satisfies the Richardson equations \eqref{Equations}, with the corresponding eigenvalues given by
\begin{eqnarray}
E\left( \mu,U,G,M,\{ \nu_{\bf k} \}\right) &=& \sum_{\mathbf{k}}\tilde{\epsilon
}_{\mathbf{k}}\nu_{\mathbf{k}} +2\sum_{\alpha}x_{\alpha
}+\frac{5U}{2}M\nonumber \\&&+\frac{U}{4}\sum_{\mathbf{k}}%
\nu_{\mathbf{k}}\left(  \nu_{\mathbf{k}}-1\right)  \text{,}
\label{Exactenergies}
\end{eqnarray}
and the number of fermions $N$ given by
\begin{equation}
N = 2M + \sum_{\bf k} \nu_{\bf k}.
\end{equation}

Indeed, as the total number of particles is conserved, one can completely eliminate the chemical potential $\mu$ and write the eigenenergies as a function of the number of fermions as 
\begin{eqnarray}
E\left( N,U,G,M,\{ \nu_{\bf k} \}\right) &=& \sum_{\mathbf{k}}\epsilon
_{\mathbf{k}}\nu_{\mathbf{k}} +2\sum_{\alpha}x'_{\alpha
}+\frac{5U}{2}M\nonumber \\&&+\frac{U}{4}\sum_{\mathbf{k}}%
\nu_{\mathbf{k}}\left(  \nu_{\mathbf{k}}-1\right)  \text{,}
\label{ExactenergiesN}
\end{eqnarray}
where $\{x'_{\alpha}\}$ are the roots of the Richardson Eqns. \eqref{Equations} with $\tilde{\epsilon}_{\bf k}$ replaced by $\epsilon_{\bf k}$.

Building on Richardson’s exact solution of the pairing Hamiltonian \cite{Richardson1963,Richardson64} and the subsequent formulation of the integrable Gaudin magnet \cite{Gaudin1976}, these models were later unified within the framework of the Richardson-Gaudin (RG) integrable systems \cite{Class2001,Colloquium2004, Ortiz2005}. The Hamiltonian in Eq. \eqref{H(U=0)} corresponds to a particular instance of the rational RG class. It can be obtained as a linear combination of the complete set of integrals of motion
\begin{equation}
R_{\mathbf{k}}=\tau_{\mathbf{k}}^{z}-2G\sum_{\mathbf{k}^{\prime}\left(
\neq\mathbf{k}\right)  }\frac{1}{\epsilon_{\mathbf{k}}-\epsilon_{\mathbf{k}%
^{\prime}}}\left[  \frac{1}{2}\left(  \tau_{\mathbf{k}}^{+}\tau_{\mathbf{k}%
^{\prime}}^{-}+\tau_{\mathbf{k}^{\prime}}^{+}\tau_{\mathbf{k}}^{-}\right)
+\tau_{\mathbf{k}}^{z}\tau_{\mathbf{k}^{\prime}}^{z}\right].
\end{equation}

Of particular interest are the hyperbolic RG models, which describe triplet pairing in spinless fermionic systems \cite{Ibanez2009, Rombouts2010}. In Appendix ~\ref{AppendixE}, we outline how our non-Fermi liquid model can be generalized by means of the hyperbolic RG family to a topological spinless fermion superfluid in $d=2$. Incorporating the spin degree of freedom necessitates extending the $su(2)$ hyperbolic model to the $so(5)$ algebra. The $so(5)$ RG topological superfluid model for spinful fermionic systems has been studied in Ref. \cite{Holdhusen2021}. An extension of this model to include non-Fermi liquid behavior is deferred to future work.

\subsection{Macroscopic Degeneracy}
\label{Macroscopic Degeneracy}

It was proved above that the energy of an arbitrary eigenstate, Eq. \eqref{Exactenergies}, of the model is entirely determined by the quantum numbers $\{ \nu_{\bf{k}}\}$ and $M$, and is independent of $\{ s_{\bf{k}}^{z} \}$ and $\{ T_{\bf{k}} \}$. This yields a four-fold degeneracy when $\nu_{\bf{k}}=1$ and a five-fold degeneracy when $\nu_{\bf{k}}=2$ (see Table \ref{table1}). Consequently, the degeneracy associated with a many-body eigenstate is given by  
\begin{equation}
\mbox{Degeneracy} = 4^{N_1} 5^{N_2},
\label{eq:degeneracy}
\end{equation}  
where \( N_1 \) is the number of momentum pairs $(\bf{k}, -\bf{k})$ present in the many-body state with $\nu_{\bf{k}}=1$, and \( N_2 \) with $\nu_{\bf{k}}=2$. Hence, we conclude that the Hamiltonian $H$ exhibits macroscopic degeneracy in its spectrum. 

\subsection{Broken $\mathbb{Z}_2$  Symmetry}
\label{Emergent symmetry}

In the non-interacting limit, $ U = G =0$, the system, which realizes a conventional Fermi liquid, exhibits an extra \(\mathbb{Z}_2\) symmetry \cite{AndersonHaldane2001} that we next show is implemented by \( \mathcal{K}_{\sigma} \). This symmetry reflects the invariance of the Fermi liquid under spin-resolved particle-hole conjugation. However, once interactions are introduced, as in the Hubbard \cite{Hubbard1963} or HK models, the spin-resolved particle-hole symmetry is explicitly broken. What remains is only the total particle-hole symmetry, implemented by $\mathcal{K}_{ph}$.

In our model, a similar symmetry structure emerges in the interacting Hamiltonian $H_{\sf n}$. The interaction term $H_{\sf U}$, which couples fermions of opposite spin at the same momentum, explicitly breaks the spin-resolved particle-hole symmetry implemented by $\mathcal{K}_{\sigma}$, leaving \( H_{\sf n} \) symmetric only under the global particle-hole transformation \( \mathcal{K}_{ph} \). One can prove that $ \mathcal{K}_{\sigma} $ acts on $ H_{\sf n} $ as 
\begin{eqnarray}
\mathcal{K}_{\sigma} H_{\sf n} \mathcal{K}_{\sigma}^{\dagger} &=& -\mu V + \sum_{\bf k}\tilde{\epsilon}_{\bf k}\hat{N}_{\bf k} + U\sum_{\bf k} \mathcal{O}_{\bf k}^{(\sigma)} \nonumber \\ &&+2\mu\sum_{\bf k}(n_{\bf k \sigma}+n_{\bf -k \sigma}) ,
\label{Asymm}
\end{eqnarray}
where 
\begin{equation}
\mathcal{O}_{\bf k}^{(\sigma)} =    A_{\bf k}^{+}A_{\bf k}^{-} 
+ \frac{1}{2}\left(A_{\bf k}^{z} +1 \right) \left(2A_{\bf k}^{z} +1 \right),
\label{Opasymm}
\end{equation}
with operators $A_{\bf k}^{+}=c_{\bf{-k+K}, \sigma}c_{\bf{-k}\bar{\sigma}}^{\dagger}+c_{\bf{k -K }, \sigma}c_{\bf{k}, \bar{\sigma}}^{\dagger}=(A_{\bf k}^{-})^{\dagger}$, $A_{\bf k}^{z} = \frac{1}{2}\left( n_{\bf{k}\bar{\sigma}} + n_{\bf{-k}\bar{\sigma}}- n_{\bf{-k+K} ,\sigma}-n_{\bf{k-K}, \sigma}\right)$ which satisfy the commutation relations of an $su(2)$ algebra  
\begin{equation}
\left[ A_{\bf{k}}^+, A_{\bf{k}}^- \right] = 2 A_{\bf{k}}^z ; \quad \left[ A_{\bf{k}}^z, A_{\bf{k}}^{\pm} \right] = \pm A_{\bf{k}}^{\pm} .
\end{equation}

This explicit symmetry-breaking has significant consequences. As demonstrated in the next section, for arbitrarily small but positive \( U\), the system exhibits \textit{multiple many-body Fermi surfaces} at zero temperature. The interacting ground states are not adiabatically connected to the non-interacting one, and their structure depends nontrivially on the strength of the interaction. These many-body Fermi surfaces signal a breakdown of the \textit{adiabatic continuity} assumption at the heart of Landau Fermi liquid theory, which posits that the low-energy excitations of an interacting system are smoothly connected to those of a non-interacting Fermi gas. Consequently, \textit{Luttinger's theorem} \cite{Luttinger1960}, which equates the volume enclosed by the Fermi surface to the total particle density, no longer holds. The appearance of such surfaces, each associated to a pole of the retarded Green's function (see Section \ref{Retarded Green's Function}) indicates the emergence of non-standard \textit{fractionalized} quasiparticles, whose properties are shaped by the system's strongly correlated dynamics \cite{Balram2015}.

\section{Quantum Phase Diagram of $H_{\sf n}$}
\label{Section3}

We first focus on the model in the absence of pairing interactions, i.e., $H_{\sf n}$, and specialize to the case of repulsive interactions (\( U > 0 \)) at zero temperature in the thermodynamic limit $L_1=L_2 \dots =L_d=L\to \infty$. We shall assume the dispersion relation to be of the tight-binding form $\epsilon_{\bf{k}}= -2t\sum_{i=1}^{d}\cos(k_i)$, where $t$ is the hopping matrix element between nearest-neighbor sites. Henceforth, we shall set $2t=1$. This is equivalent to measuring $U$ and $\mu$ in units of $2t$.

In order to calculate the ground state of $H_{\sf n}$, we begin by noticing that since $H_{\sf n} = \sum_{\bf k} H_{{\sf n}, \bf k}$, where
\begin{equation}
H_{{\sf n}, \bf k} = \tilde{\epsilon}_{\bf k} \hat{N}_{\bf k} + U \left( \tau_{\bf k}^{+}\tau_{\bf k}^{-} + \frac{1}{4} \hat{N}_{\bf k} (\hat{N}_{\bf k}-1) \right)
\end{equation}
consists of operators which act non-trivially only on the subspace $\mathcal{H}_{\bf k}$, an arbitrary (unnormalized) eigenstate $|\Psi\rangle$ may be written as the product state 
\begin{equation}
| \Psi \rangle = \prod_{\bf k} \left( \tau_{\bf k}^+ \right)^{M_{\bf k}} | \lambda_{\bf k} \rangle ,
\end{equation}
with eigenvalue (see Table \ref{table1})
\begin{equation}
E (\mu, U) = \sum_{\bf k } \tilde{E}_{\bf k} (\mu, U) .
\label{decompose}
\end{equation}
In order to obtain the ground state of $H_{\sf n}$, we perform an integer minimization of $E(\mu, U)$, fixing $\mu$ and $U$, while allowing the number of fermions to vary
\begin{eqnarray}
E_{0}(\mu,U) &=& \min_{\{0 \le M_{\bf k}+\nu_{\bf k}\le 2\}} \sum_{\bf k} \tilde E_{\bf k}(\mu,U).
\end{eqnarray}

Since the many-body energy is a sum of independent energies,  finding the minimum of $E(\mu, U)$ reduces to finding the minimum of $\tilde{E}_{\bf k}(\mu, U)$ for each value of $\bf{k}$. This procedure yields the following expressions for the ground state and ground-state energy: Consider partitioning the half-BZ (characterized by $k_1>0$) into the following five sectors 
\begin{eqnarray}
{\cal S}_{4} &=& \{{\bf k} \, | -d-\mu <  \tilde{\epsilon}_{\bf k} \leq \tilde{\epsilon}_4 \}, \nonumber \\
{\cal S}_{3} &=& \{{\bf k} \,| \, \tilde{\epsilon}_4 <  \tilde{\epsilon}_{\bf k} \leq  \tilde{\epsilon}_3 \}, \nonumber \\
{\cal S}_{2} &=& \{{\bf k} \, |\,  \tilde{\epsilon}_3 <  \tilde{\epsilon}_{\bf k} \leq \tilde{\epsilon}_2 \}, \nonumber \\
{\cal S}_{1} &=& \{{\bf k} \, |\,  \tilde{\epsilon}_2 <  \tilde{\epsilon}_{\bf k} \leq \tilde{\epsilon}_1\}, \nonumber \\
{\cal S}_{0} &=& \{{\bf k} \, | \,  \tilde{\epsilon}_1 <  \tilde{\epsilon}_{ \bf k} <  d-\mu \},
\label{k-sectors}
\end{eqnarray}
where the parameters $\tilde{\epsilon}_j$, $j=1,\cdots,4$, represent the four many-body Fermi-energies of our model
\begin{eqnarray}
\tilde{\epsilon}_4&=& \gamma\left(\frac{-5U}{2}, \mu\right) \nonumber , \, 
\tilde{\epsilon}_3= \gamma\left(-2U, \mu\right) \nonumber ,\\
\tilde{\epsilon}_2 &=& \gamma\left(\frac{-U}{2}, \mu\right) , \, 
\tilde{\epsilon}_1=\gamma\left(0, \mu\right),
\label{fermisurfaces}
\end{eqnarray}
defined in terms of the function 
\begin{eqnarray}
   \gamma(x, \mu) = \min (\max(x, -d-\mu), d-\mu) .
\end{eqnarray} 
Then, the (unnormalized) ground state of $H_{\sf n}$ is
\begin{equation}
|\Psi_0\rangle =   \prod_{\bf k \in \mathcal{S}_4} (\tau_{\bf k}^+)^2 \prod_{\bf k \in \mathcal{S}_3}  \tau_{\bf k}^+ |\lambda_0\rangle ,
\label{gsofhn}
\end{equation}
where $|\lambda_0\rangle$ are pairon vacua, i.e., states annihilated by all the $\tau_{\bf k}^-$, and characterized by the quantum numbers $\nu_{\bf k} =0$ for ${\bf k} \in {\cal S}_4 \cup {\cal S}_0$, $\nu_{\bf k} =1$ for ${\bf k} \in {\cal S}_3 \cup {\cal S}_1$ and $\nu_{\bf k} =2$ for ${\bf k} \in {\cal S}_2$. 
The ground-state energy is given by 
\begin{eqnarray}
E_{0}(\mu, U) &=& \sum_{\bf k \in {\cal S}_{4}} \left( 4\tilde{\epsilon}_{\bf{k}}+ 5U  \right) + \sum_{\bf k \in {\cal S}_{3}} \left( 3\tilde{\epsilon}_{\bf{k}}+ \frac{5U}{2}  \right) \nonumber \\
&& + \sum_{\bf k \in {\cal S}_{2}} \left( 2\tilde{\epsilon}_{\bf{k}}+ \frac{U}{2}  \right) + \sum_{\bf k \in {\cal S}_{1}}  \tilde{\epsilon}_{\bf{k}} .
\end{eqnarray}

One can subsequently proceed to calculate the number of fermions in the ground state by differentiating $E_{0}(\mu,U)$ w.r.t. $\mu$, i.e., $-\partial E_{0}(\mu,U)/\partial \mu= N $ . (By inverting this relation at fixed $U$, one obtains $\mu$ as a function of $N$.) Finally, one can calculate the ground-state energy density $\epsilon_0(\rho_F, U)$ as a function of the fermion number density $\rho_F \equiv \frac{N}{V}$ and $U$ as 
\begin{eqnarray}
\epsilon_{0}(\rho_F, U) &=&  \frac{1}{V}(E_0(\mu, U)+\mu N ) \nonumber \\
&=&\frac{1}{V}\sum_{\bf k \in {\cal S}_{4}} \left( 4\epsilon_{\bf{k}}+ 5U  \right) + \frac{1}{V}\sum_{\bf k \in {\cal S}_{3}} \left( 3\epsilon_{\bf{k}}+ \frac{5U}{2}  \right) \nonumber \\
&&+  \frac{1}{V}\sum_{\bf k \in {\cal S}_{2}} \left( 2\epsilon_{\bf{k}}+ \frac{U}{2}  \right) + \frac{1}{V}\sum_{\bf k \in {\cal S}_{1}}  \epsilon_{\bf{k}} .
\label{energydensity}
\end{eqnarray}

\begin{figure}[bth]
    \centering
\includegraphics[width=0.45\textwidth]{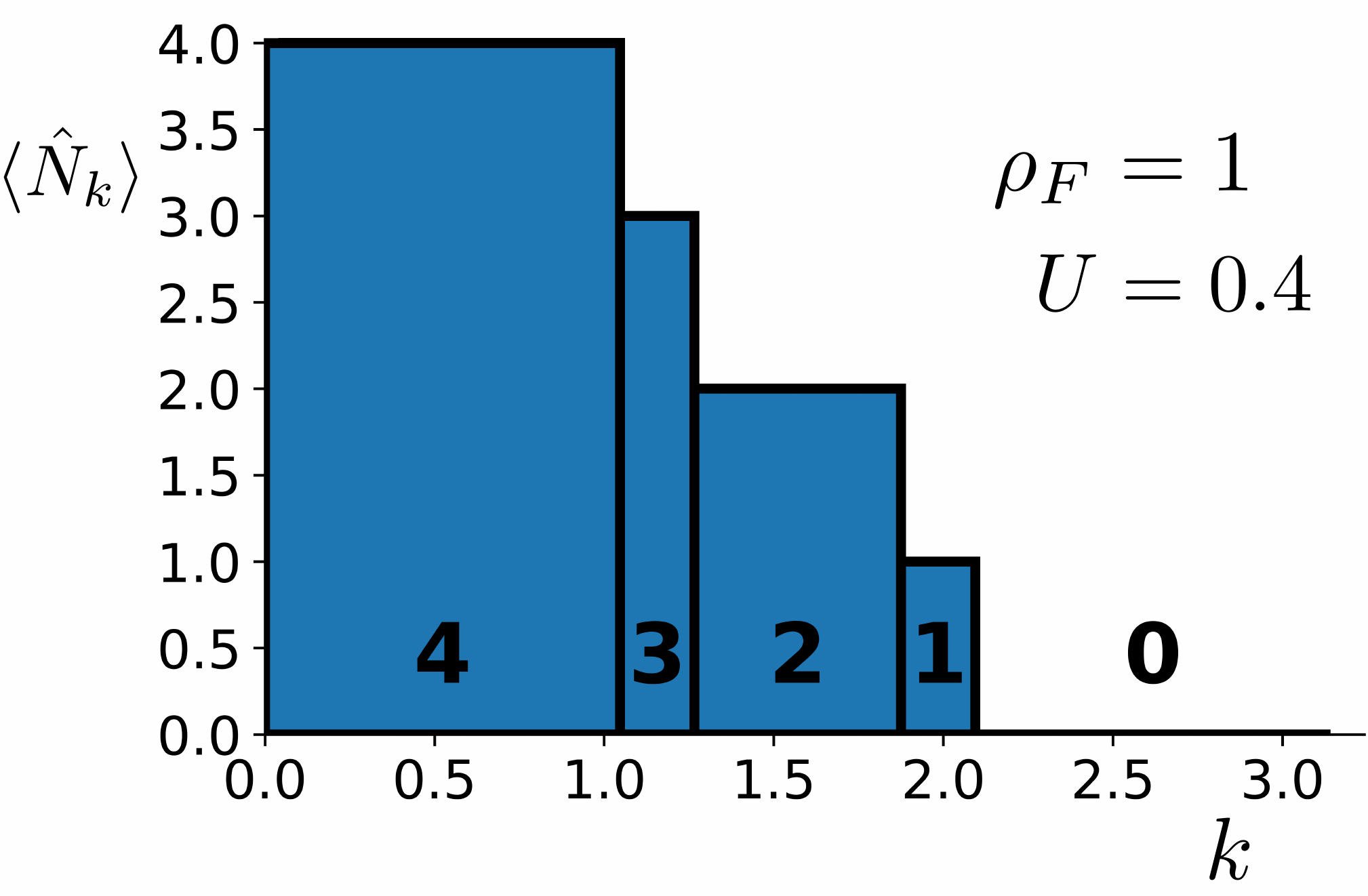}
    \caption{Occupation numbers for the ground state of phase \textquotedblleft(4,3,2,1,0)\textquotedblright of $H_{\sf n} $ in one spatial dimension.}
    \label{fig:nkvsk1d}
\end{figure}

We now investigate the quantum phase diagram of the model (with $G = 0$) in several spatial dimensions $d$. As we vary the strength of the repulsion $U$ and the fermion density $\rho_F$ (or equivalently, the chemical potential \( \mu \)), we observe a total of ten metallic phases and three Mott-insulating phases.

A convenient way of labeling phases is to list the sectors, $\mathcal{S}_j$, $j=0,\cdots,4$,  that are not empty in the ground-state wavefunction of the corresponding phase. This is equivalent to labeling each quantum phase by the occupancy patterns of energy levels in the ground state. The occupation number for a given density $\rho_F$ and repulsion strength $U$ is defined as
\begin{equation}
\langle \hat{N}_{\bf k}\rangle=\frac{\langle \Psi_0 | \hat{N}_{\bf k} | \Psi_0 \rangle}{\langle \Psi_0|\Psi_0\rangle} . 
\end{equation}

For example, the phase \textquotedblleft (4,3,2,1)\textquotedblright is characterized by $ {\cal S}_{0} = \varnothing$ (i.e., no ${\bf k}$'s in the set ${\cal S}_{0}$), while the phase \textquotedblleft(3,2)\textquotedblright by ${\cal S}_{4},  {\cal S}_{1}, {\cal S}_{0} = \varnothing$.  This is illustrated in Fig. ~\ref{fig:nkvsk1d}, in which we show the occupation numbers $\langle \hat{N}_{\bf k}\rangle$ corresponding to the phase \textquotedblleft(4,3,2,1,0)\textquotedblright in one spatial dimension.

Given the many-body Fermi energies (see Eq.\eqref{fermisurfaces}), the phase transition points can be determined as functions of $U$ and $\mu$. For example, consider the phase transition between phases \textquotedblleft(4,3,2,1,0)\textquotedblright and \textquotedblleft(3,2,1,0)\textquotedblright. At the phase transition, the location of the many-body Fermi surface separating $\mathcal{S}_4$ and $\mathcal{S}_3$ must be at $\bf k = 0$. This yields the location of the phase transition at $\mu= -d+\frac{5}{2}U$. Similar calculations provide analytic expressions for all the phase transitions, and are listed in Tables \ref{tablemm} and \ref{tablemi}. 
The order of the phase transitions is determined by the non-analyticities of the ground-state energy density $\epsilon_0(\rho_F, U)$. 

\begin{table}[htb]
\[
\begin{array}{|c|c|c|c|c|}
\hline
 \text{Phase at $\mu_{c-}$} & \text{Phase at $\mu_{c+}$}& \mu_c & \text{Bounds for $U$} \\
\hline
 (1,0) & (2,1,0) & -d +\frac{U}{2} & 0<U<4d\\
 (2,1,0)& (2,1) & d & d<U<4d\\
 (2,1) & (3,2,1) &-d + 2U&d<U<\frac{4d}{3}\\
 (3,2, 1) & (3,2) & d + \frac{U}{2}& d<U<\frac{4d}{3}\\
 (3,2) & (4,3,2) &  -d + \frac{5U}{2}& d<U<4d\\
 (4,3,2) & (4,3) &  d + 2U&0<U<4d\\
 (2,1,0) & (3,2,1,0) & -d + 2U&0<U<d\\
 (3,2,1,0) & (3,2,1) &  d&\frac{4d}{5}<U<d\\
 (3,2,1) & (4,3,2,1) &  -d + \frac{5U}{2}&\frac{4d}{5}<U<d\\
 (4,3,2,1) & (4,3,2) &  d + \frac{U}{2}&0<U<d\\
 (3,2,1,0) & (4,3,2,1,0) & -d +\frac{5U}{2}&0<U<\frac{4d}{5}\\
 (4,3,2,1,0) & (4,3,2,1) & d&0<U<\frac{4d}{5}\\
\hline
\end{array}
\]
\caption{Metal-metal transitions in the quantum phase diagram of $H_{\sf n}$ for an arbitrary number of spatial dimensions $d$. In this table $\mu_{c\pm} = \mu_c \pm \delta$, where $\mu_c$ is the value of $\mu$ at the phase transition and $\delta$ is a small positive number.} 
\label{tablemm}
\end{table}

\begin{table}[htb]
\[
\begin{array}{|c|c|c|c|c|}
\hline
 \text{Phase at $\mu_{c-}$} & \text{Phase at $\mu_{c+}$} & \mu_c & \text{Bounds for $U$} \\
\hline
  (0)& (1, 0) &  -d &U>0\\
  (1,0)& (1) &  d & U>4d\\
  (1) & (2,1) & -d + \frac{U}{2}&U>4d\\
  (2, 1) & (2) & d + \frac{U}{2}&U>\frac{4d}{3}\\
  (2) & (3,2) & -d + 2U&U>\frac{4d}{3}\\
  (3,2) & (3) &  d+ 2U& U>4d\\
  (3) & (4,3) &  -d + \frac{5U}{2}&U>4d\\
  (4,3) & (4) & d + \frac{5U}{2}&U>0\\
\hline
\end{array}
\]
\caption{Metal-insulator transitions in the zero temperature phase diagram of $H_{\sf n}$ for an arbitrary number of spatial dimensions $d$.In this table $\mu_{c\pm} = \mu_c \pm \delta$, where $\mu_c$ is the value of $\mu$ at the phase transition and $\delta$ is a small positive number. } 
\label{tablemi}
\end{table}

Similar considerations allow us to identify three Mott-insulating phases. These phases always appear as lines of first-order transitions in the $\rho_F$-$U$ phase diagram. They are found at densities $\rho_F=\frac{1}{2},1,\frac{3}{2}$, for repulsive interaction strengths $U>U_{c}$, where
\begin{equation}
U_{c}(\rho_F) = \frac{4d}{3} \left( 1+4|\rho_F -1| \right)  . 
\label{criticalu}
\end{equation} 

The conduction properties of each phase are determined by analyzing the {\it charge gap}, namely the discontinuity of the chemical potential $\mu(\rho_F)$ at zero temperature
\begin{equation}
\Delta \mu (\rho_F)=\lim_{\delta \rightarrow 0^+} \left ( \, \mu\big|_{\rho_F +\delta}- \mu\big|_{\rho_F-\delta}
\, \right ) .
\end{equation}
It can be proved that if \( \rho_F \neq \frac{1}{2}, 1, \frac{3}{2} \), then \( \Delta \mu =0 \) in all spatial dimensions $d$. Hence, away from quarter-filling, half-filling, and three-quarters filling, the system is always a metal. 

Let us now consider the three special fillings $\rho_F = \frac{1}{2}, 1, \frac{3}{2}$. For these special fillings,
\begin{eqnarray}
   \Delta \mu (\rho_F)  =
\begin{cases}
\text{0}, & \text{if $ U<U_c(\rho_F)$}, \\
\text{$2d \left( \frac{U}{U_c(\rho_F)} -1\right)$}, & \text{if $U>U_c(\rho_F)$}.
\end{cases}
\end{eqnarray}
The emergence of a finite charge gap indicates an interaction-driven insulating phase, which motivates identifying phases \textquotedblleft(1)\textquotedblright, \textquotedblleft(2)\textquotedblright, and \textquotedblleft(3)\textquotedblright as Mott-insulating phases. 

The following sections discuss features of the quantum phase diagram that depend on spatial dimensionality and establish a mathematical correspondence between the HK model and our model. Additional details regarding the HK model are provided in Appendix~\ref{AppendixHK}. 

\subsection{One and Two Spatial Dimensions}
\label{One Dimension}

\begin{figure*}[t]
    \centering  \hspace*{-0.74cm}
    \includegraphics[width=0.46\textwidth]{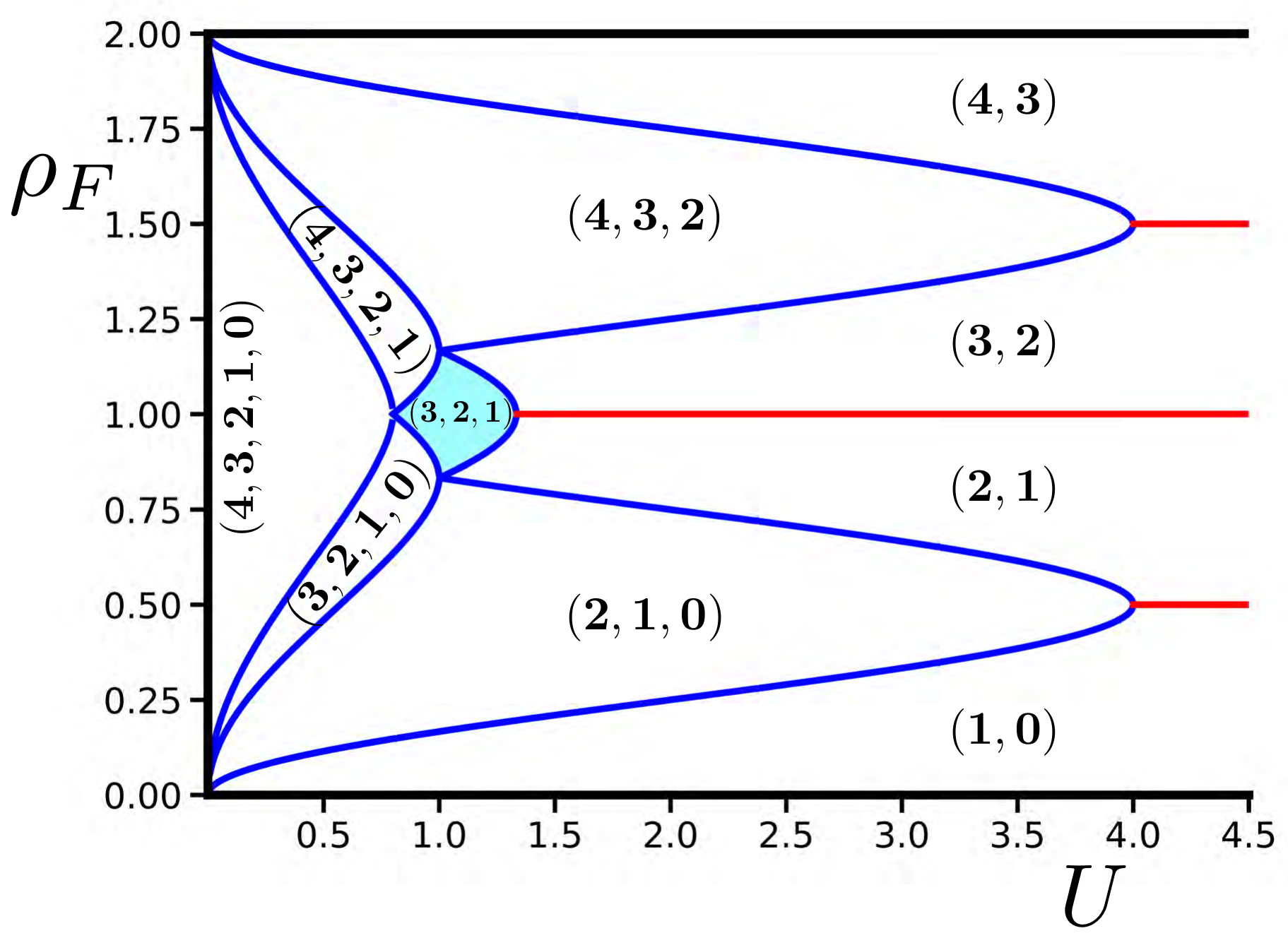} \hspace*{0.3cm}
    \includegraphics[width=0.46\textwidth]{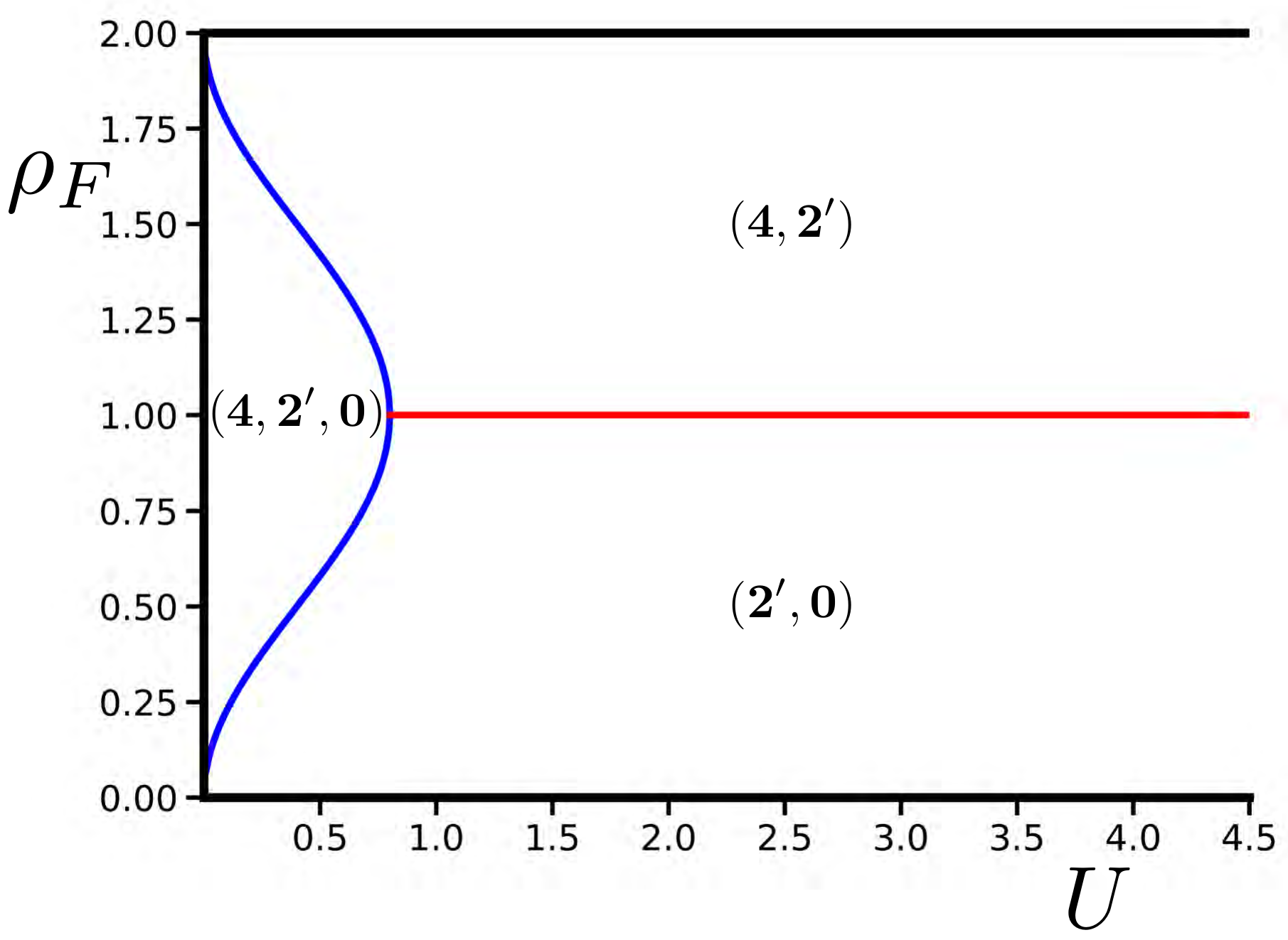} \hspace*{-0.7cm}
    \includegraphics[width=0.45\textwidth]{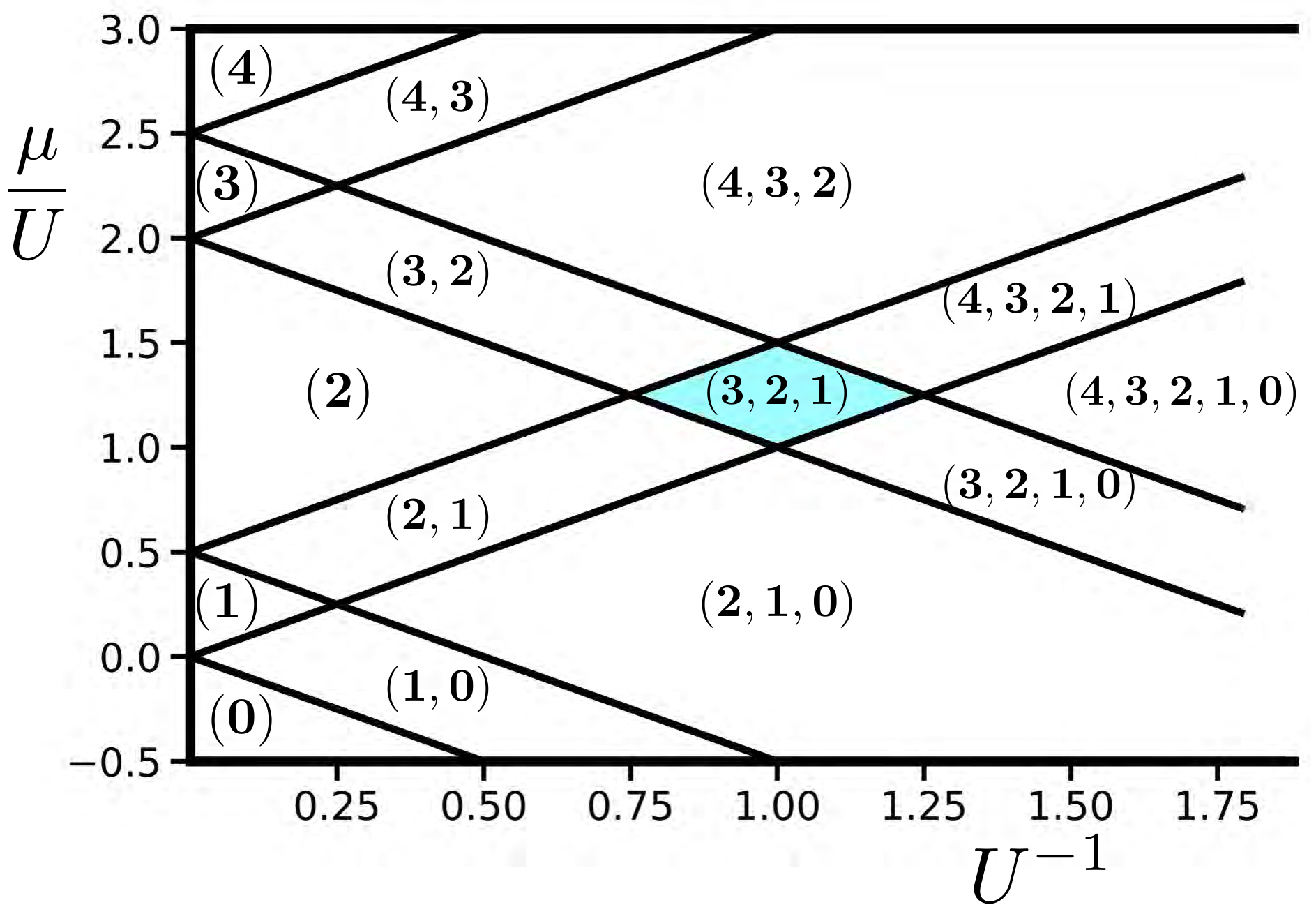} \hspace*{0.5cm}
    \includegraphics[width=0.45\textwidth]{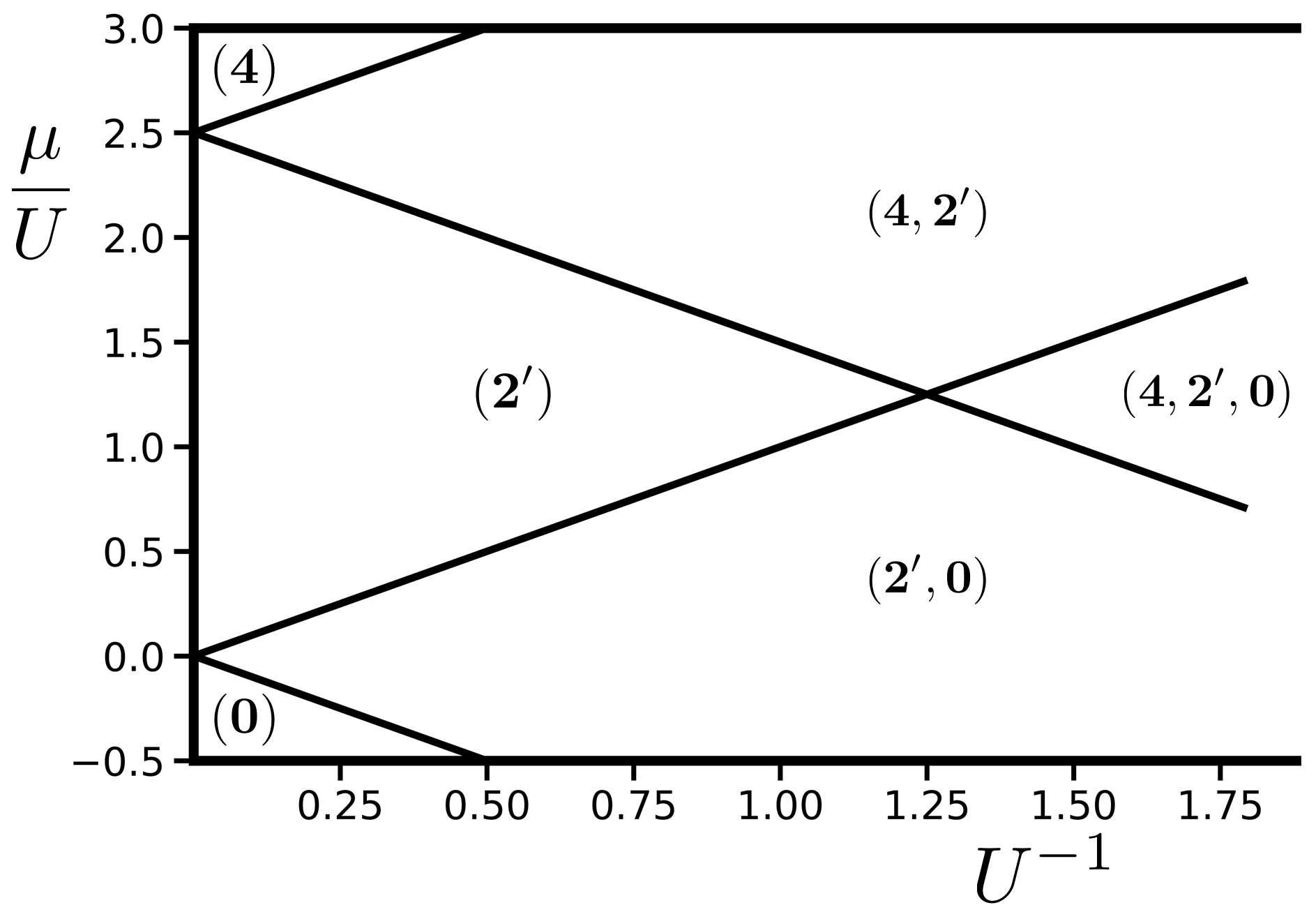}
    \caption{Quantum phase diagram for our model (left) in $d=1$ contrasted with the HK model (right). (Refer to Appendix \ref{AppendixHK} for a detailed discussion of the quantum phase diagram of the HK model.) Band insulators (BI) are found at full filling ($\rho_F =2$) and empty filling ($\rho_F =0$). Mott insulators (MI) are found at $\rho_F = 1/2$, $1$ and $3/2$ for $U>U_c(\rho_F)$. The blue lines represent second order and the red lines represent first order transitions. The red lines coincide with the Mott-insulating phases. }
    \label{fig:phases}
\end{figure*}

The quantum phase diagrams of both our model and the HK model in one spatial dimension ($d=1$) are shown in Fig.~\ref{fig:phases}.  We find that the metal-metal transitions are second-order phase transitions and are shown in \textcolor{blue}{blue}, while the first-order phase transitions (coinciding with the Mott-insulating phases) are shown in \textcolor{red}{red} in Fig. ~\ref{fig:phases}. 

\begin{figure*}[t]
    \centering    
    \hspace*{-0.68cm} \includegraphics[width=0.46\textwidth]{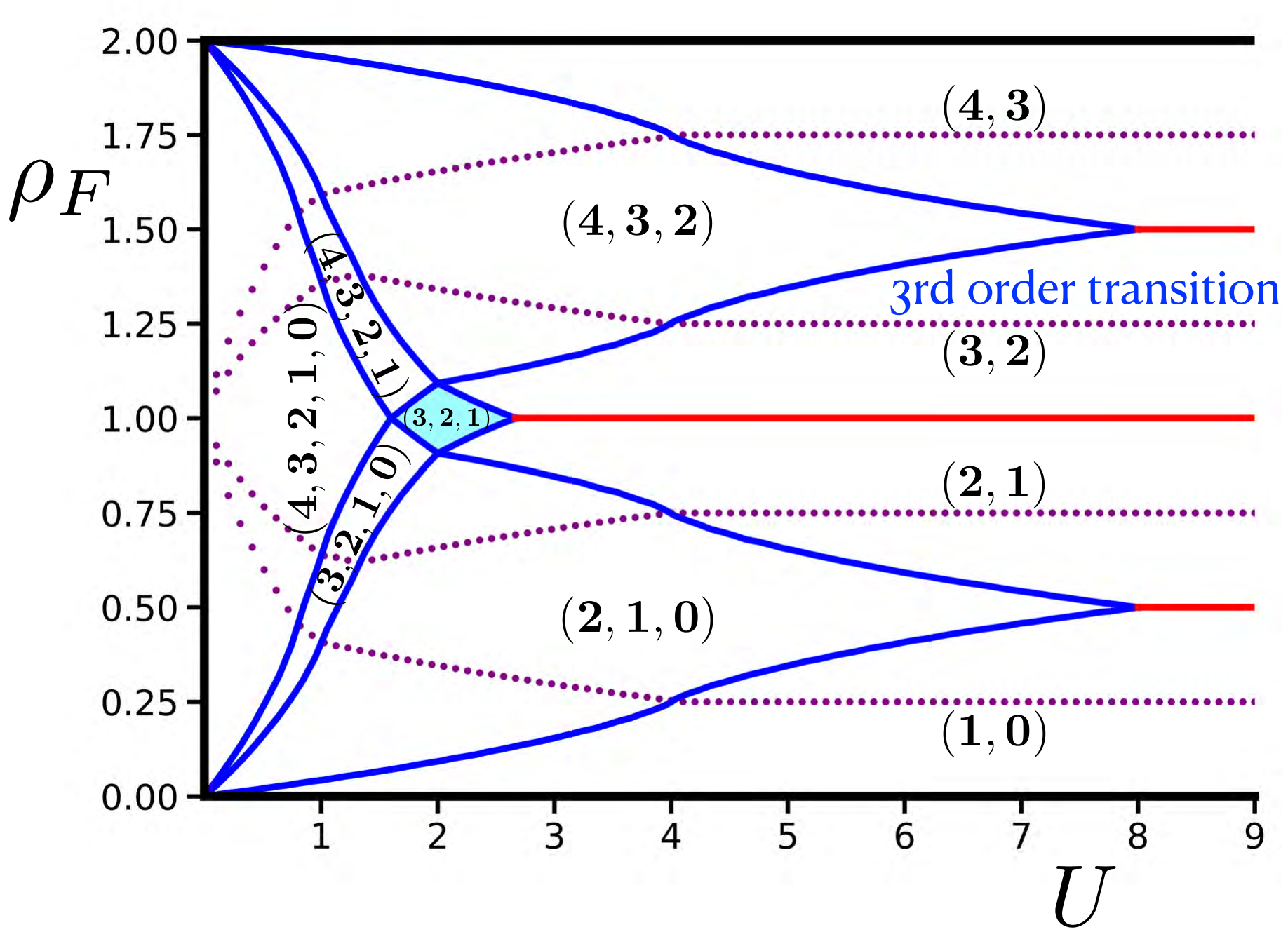}
    \hspace*{0.1cm} \includegraphics[width=0.46\textwidth]{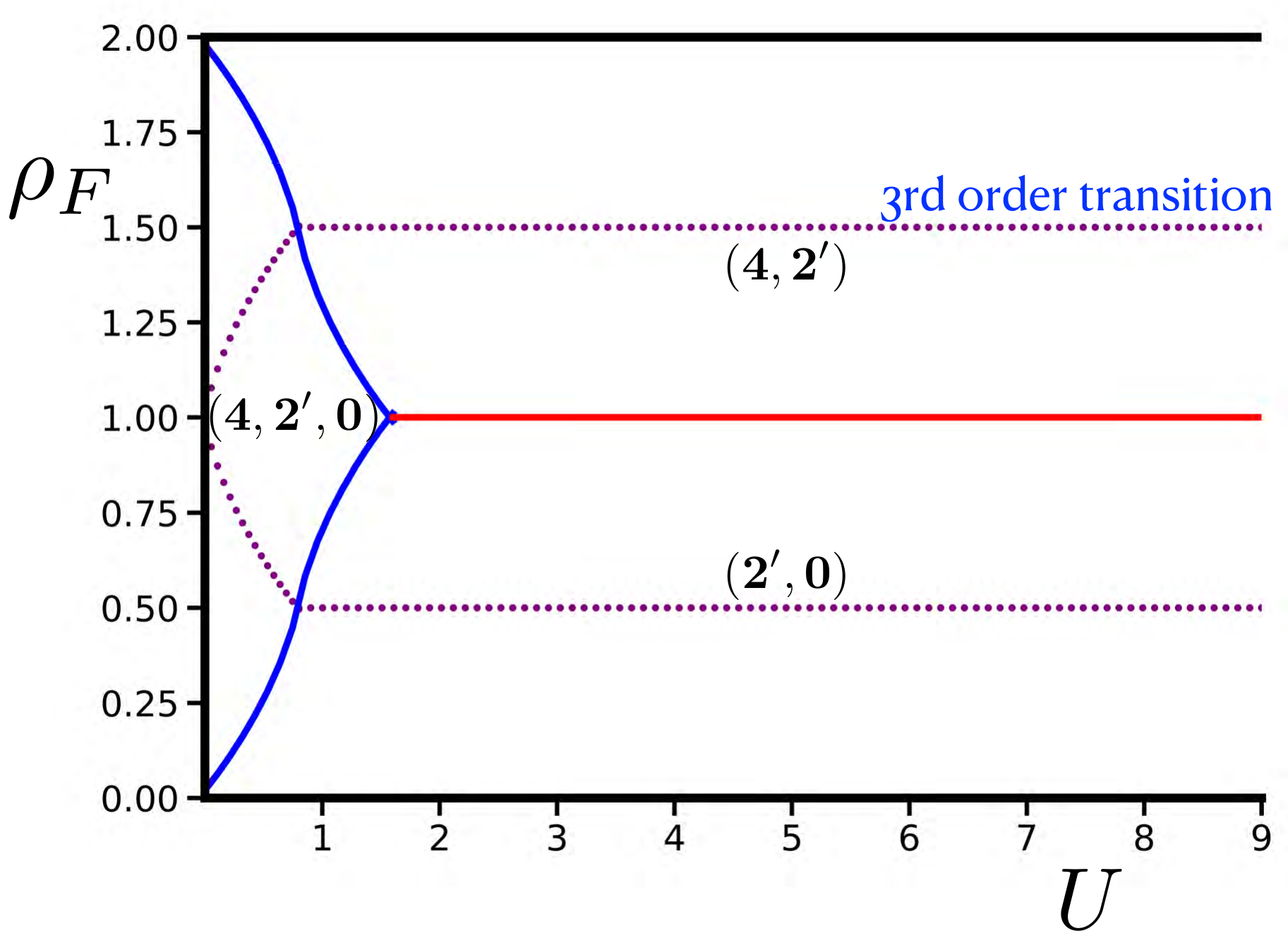} \hspace*{-0.7cm}
    \includegraphics[width=0.45\textwidth]{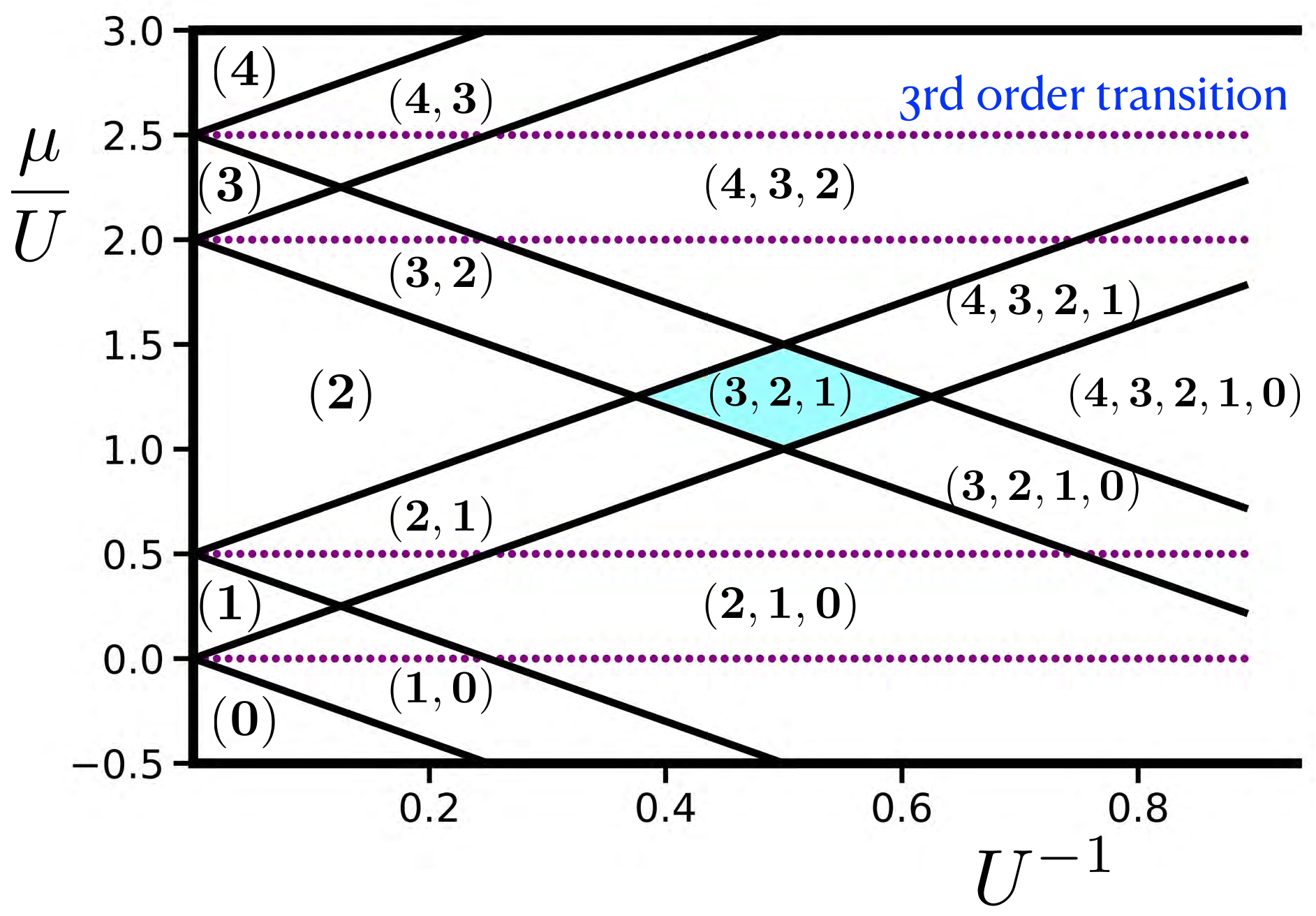}\hspace*{0.5cm}
    \includegraphics[width=0.45\textwidth]{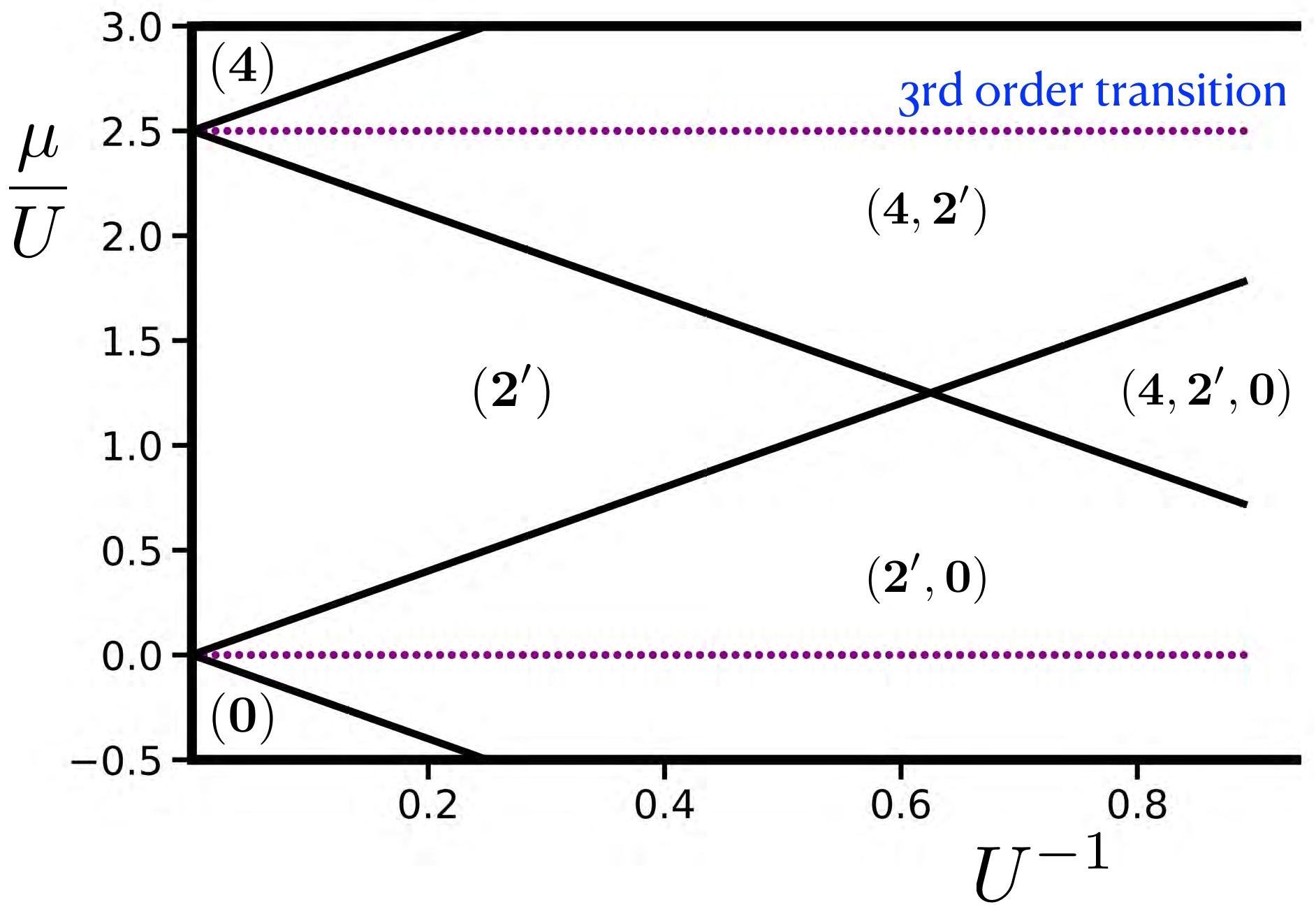}
    \caption{Quantum phase diagram for our model (left) in $d=2$  contrasted with the HK model (right). Refer to Appendix \ref{AppendixHK} for a detailed discussion of the quantum phase diagram of the HK model.) Band insulators (BI) are found at full filling ($\rho_F=2$) and empty filling ($\rho_F =0$). Mott insulators (MI) are found at $\rho_F = 1/2$, $1$ and $3/2$ for $U>U_c(\rho_F)$. The purple dotted lines indicate the third order transitions driven by the van-Hove singularity in $d=2$. The blue lines represent second order and the red lines represent first order transitions. The red lines coincide with the Mott-insulating phases.}
\label{fig:phasesmu-rho}
\end{figure*}

\begin{figure*}[t]
    \centering    \hspace*{-0.7cm}
\includegraphics[width=0.45\textwidth]{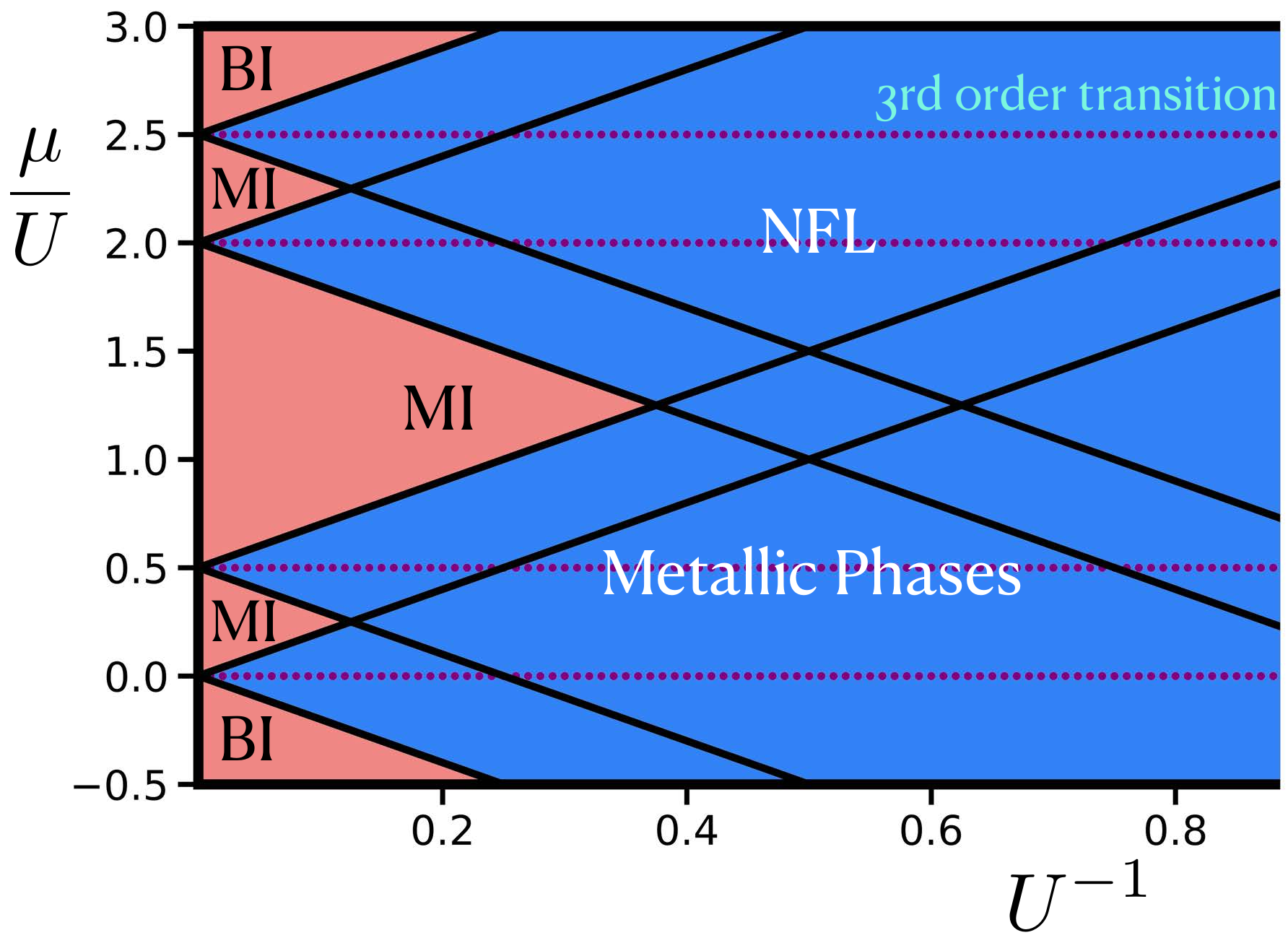} \hspace*{0.5cm}
\includegraphics[width=0.45\textwidth]{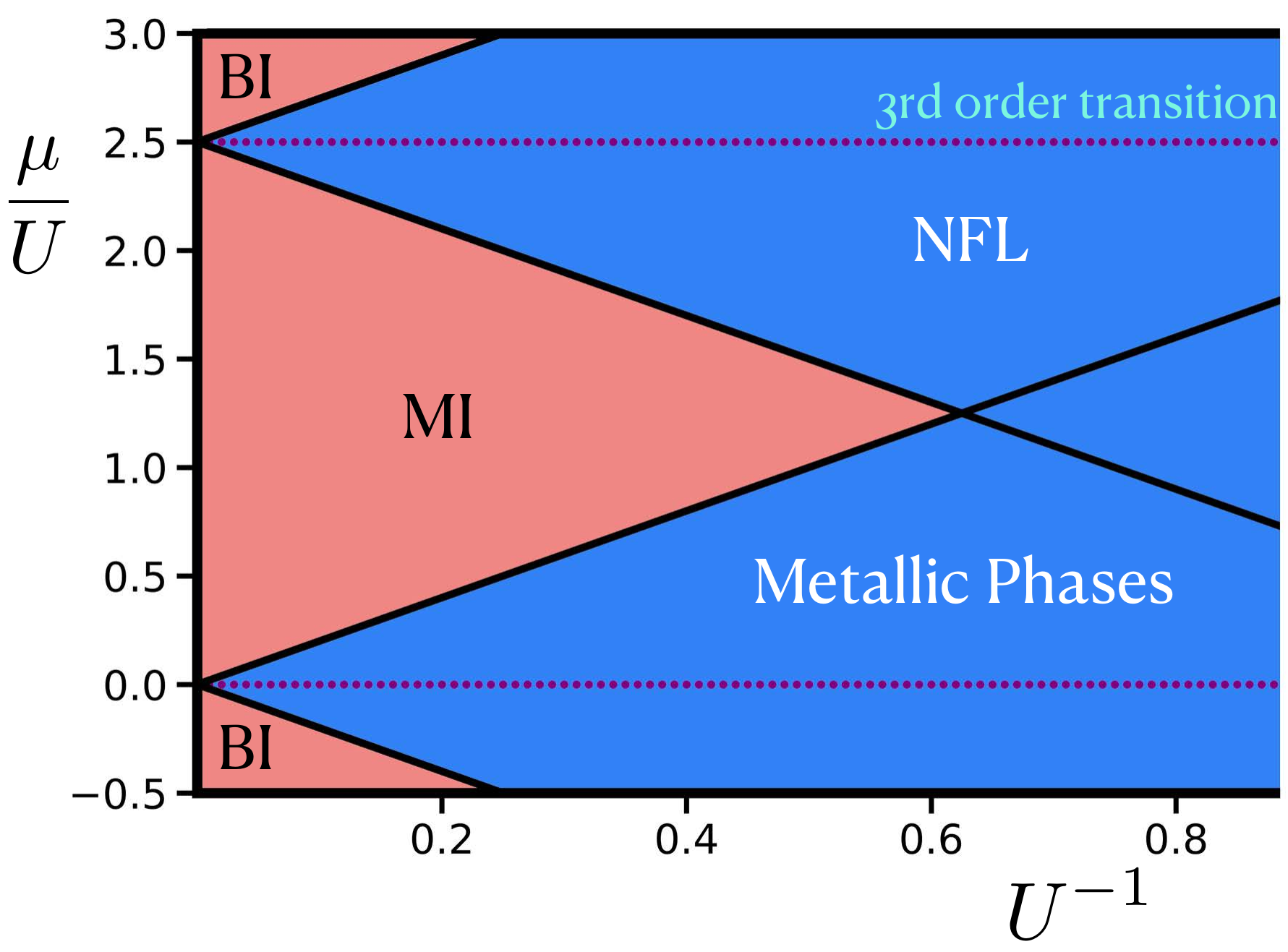} 
    \caption{Quantum phase diagram for our model (left) in $d=2$ contrasted with the HK model (right). Band insulators (BI) are found at full filling ($\rho_F =2$) and empty filling ($\rho_F =0$). Mott insulators (MI) are found at $\rho_F = 1/2$, $1$ and $3/2$ for $U>U_c(\rho_F)$. The purple dotted lines indicate the third order transitions driven by the van-Hove singularity in $d=2$.}
\label{fig:phasesmu}
\end{figure*}

\begin{figure*}[t]
    \centering
\includegraphics[width=0.95\textwidth]{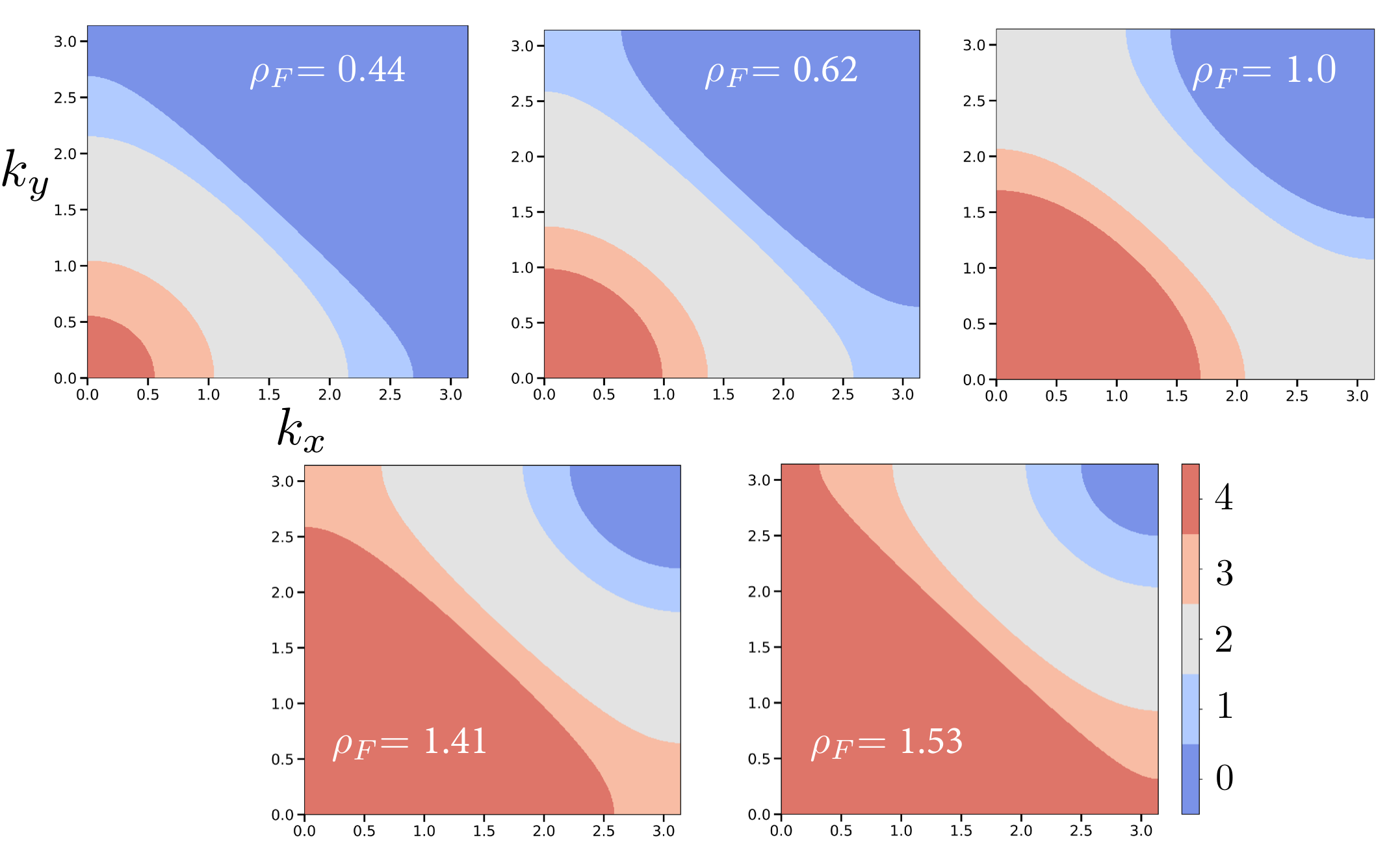}
    \caption{Occupation numbers for the ground state of phase \textquotedblleft(4,3,2,1,0)\textquotedblright of $H_{\sf n}$, as a function of density $\rho_F$,  in $d=2$ and $U=0.7$.}
    \label{fig:fs43210}
\end{figure*}

\begin{figure*}[t]
    \centering
    \includegraphics[width=0.95\textwidth]{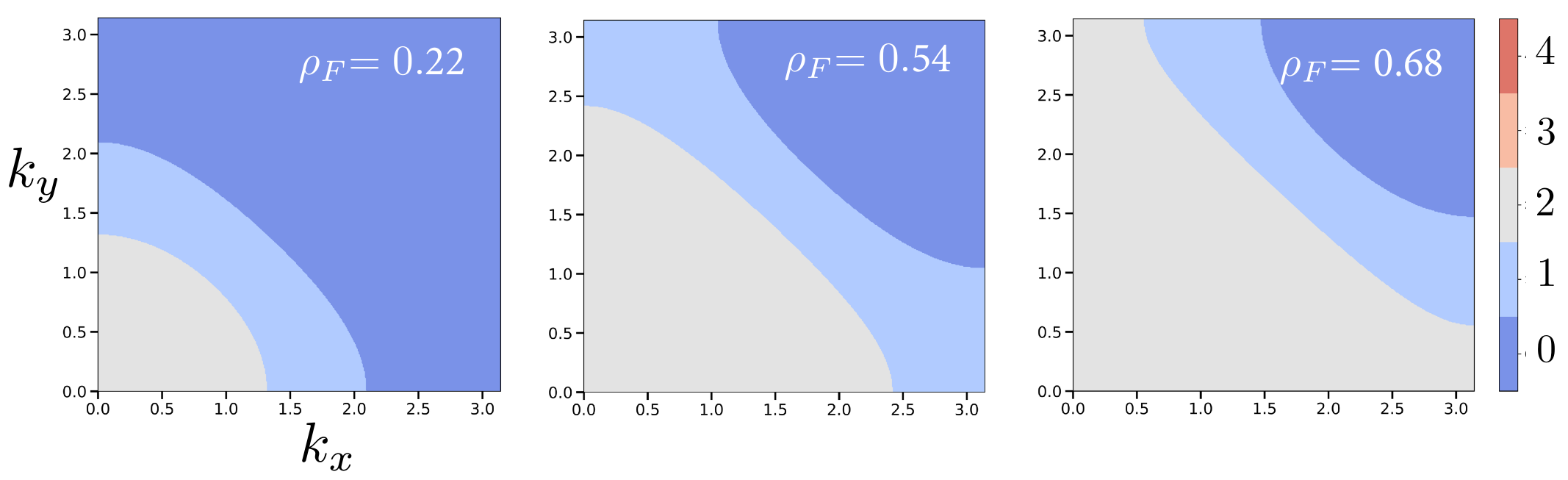}
    \caption{Occupation numbers for the ground state of phase \textquotedblleft(2,1,0)\textquotedblright of $H_{\sf n}$, as a function of density $\rho_F$,  in $d=2$ and $U=1.5$. }
    \label{fig:fs210}
\end{figure*}

In two spatial dimensions, although the positions of the phase transitions are modified, as shown in Figs.~\ref{fig:phasesmu-rho} and ~\ref{fig:phasesmu}, the order of the phase transitions remains the \textit{same} as in $d=1$. Interestingly, in $d=2$, we find four lines of \textit{third-order} phase transitions criss-crossing the phase diagram, corresponding to many-body Lifshitz transitions. Each of these transitions occurs whenever a many-body Fermi surface crosses the half-filling line in the half-BZ, defined by $\epsilon_{\bf{k}} = 0$, as shown in Figs. \ref{fig:fs43210} and~\ref{fig:fs210}. These transitions reflect qualitative changes in the shape of the many-body Fermi surfaces, and arise from the logarithmic Van-Hove singularity in $d=2$, which signals a topological change in the Fermi surface. 


\begin{table}[htb]
\[
\begin{array}{|c|c|c|c|c|}
\hline
 & N_- & N_+ & \text{Many-body F.S. }& \text{Lifshitz Transition in $d=2$}\\
\hline
1 & 1& 0 & \tilde{\epsilon}_{{\bf{k}}_F}=\tilde{\epsilon}_1 & \mu = 0\\
2 & 2& 1 & \tilde\epsilon_{{\bf{k}}_F}=\tilde{\epsilon}_2 & \mu = \frac{U}{2}\\
3 & 3 & 2 &\tilde\epsilon_{{\bf{k}}_F}=\tilde{\epsilon}_3& \mu = 2U\\
4 & 4 & 3 & \tilde\epsilon_{{\bf{k}}_F} = \tilde{\epsilon}_4 & \mu = \frac{5U}{2}\\
\hline
\end{array}
\]
\caption{Many-body Fermi surfaces (for arbitrary $d$) and associated many-body Lifshitz transitions in $d=2$. We define $N_{\pm} = \langle \hat{N}_{\bf k} \rangle_{{\bf k}_F \pm \hat \delta_{{\bf k}_F}}$, where $\hat \delta_{{\bf k}_F}=\delta  \frac{\nabla_{\bf k} \tilde{\epsilon}_{\bf k}}{|\nabla_{\bf k} \tilde{\epsilon}_{\bf k}|} \rvert_{{\bf k = k}_F}$ for a small positive $\delta$.} 
\label{table2}
\end{table}

Since the exact many-body Fermi energies are known (see Eq.~\eqref{fermisurfaces}), the locations of the four many-body Lifshitz transitions can be calculated analytically and are listed in the last column of Table~\ref{table2}. The qualitative differences between the ground states on either side of these transitions are clearly illustrated for the phases \textquotedblleft(4,3,2,1,0)\textquotedblright\ and \textquotedblleft(2,1,0)\textquotedblright\ in Figs.~\ref{fig:fs43210} and~\ref{fig:fs210}. As evident from the figures, a Lifshitz transition separates regions of a metallic phase where the many-body Fermi surface changes from ``particle-like'' to ``hole-like.''

We emphasize that the four lines corresponding to many-body Lifshitz transitions do not intersect every thermodynamic phase. The maximum number of these lines that can cross a given phase is determined by the number of many-body Fermi surfaces present in that phase. For example, the phase \textquotedblleft(4,3,2,1,0)\textquotedblright\ is intersected by all four lines, whereas the phase \textquotedblleft(3,2,1)\textquotedblright\ is not crossed by any of them. This can be seen most clearly in the quantum phase diagram $\mu/U$  vs $U^{-1}$ in the lower panels of Fig. \ref{fig:phasesmu-rho}.

\subsection{Higher Spatial Dimensions}
\label{Higher Dimensions}


In spatial dimensions $d>2$, the same ten metallic phases and three Mott insulating phases exist. Many-body Lifshitz transitions are observed criss-crossing the phase diagram, with their number dictated by the Van Hove singularities of the free Fermi liquid corresponding to $U=0$ and $\mu=\mu_0$ ($\mu_0=0$ in $d=2$). When repulsive interactions of strength $U$ are present, each $\mu_0$ of the free Fermi liquid gives rise to four many-body Lifshitz transitions, described by
\begin{equation}
\mu - \mu_0 = \frac{5U}{2}, \, 2U, \,  \frac{U}{2}, \, 0 \ .  
\end{equation}

\subsection{Emergence of the HK ground-state energy}
\label{HKemergence}

{\it Is there a relation between the HK and our model's ground-state energies}? Given the expression for the eigenstates of $H$, Eq.~\eqref{Ansatz}, we note that once the pairing terms are included, $M_{\bf k}$ is no longer a good quantum number. In contrast, $s^z_{\bf k}$, $T_{\bf k}$, and $\nu_{\bf k}$ remain conserved. To develop a framework for understanding the emergence of superconductivity in our model, we now focus on the structure of the pairon vacua $|\lambda\rangle$. Specifically, let $P$ denote the projection operator onto the subspace spanned by these pairon vacua. We then pose the following question: \textit{What is the quantum phase diagram of $PH_{\sf n}P$?}

Let us begin by setting $M_{\bf k}=0$ in Eq.~\eqref{spenergies} to obtain the projected energy associated with the pairon vacuum $|\lambda\rangle$, which we define as 
\begin{eqnarray}
E(\mu, U, M=0) &=& \sum_{\bf k}\tilde{E}_{\bf k}(\mu, U, M=0) \nonumber \\
&=& \sum_{\bf k} \tilde{\epsilon}_{\bf k} \nu_{\bf k} + \frac{U}{4}\sum_{\bf k}\nu_{\bf k} (\nu_{\bf k}-1) .
\end{eqnarray}
Next, we fix $\mu$ and $U$, and perform an integer minimization analogous to that used in determining the ground-state energy of $H_{\sf n}$, while allowing $\nu_{\bf k}$ to vary. Since the many-body energy is again a sum of independent single-site energies, minimizing $E(\mu, U, M=0)$ once again reduces  to finding the minimum of $\tilde{E}_{\bf k}(\mu, U, M=0)$ for each momentum $\bf{k}$.  We now apply this procedure to determine the projected ground state and its energy.

Consider partitioning the half-BZ into three sectors: $\tilde {\cal S}_2={\cal S}_4 \cup {\cal S}_3 \cup {\cal S}_2, {\cal S}_1, {\cal S}_0$, Eq. \eqref{k-sectors}. Then, the projected ground states are
\begin{eqnarray}
    |\overline \Psi_0\rangle=|\lambda_0'\rangle , 
\end{eqnarray}
where $|\lambda_0'\rangle$ are  pairon vacua, i.e., states annihilated by all the $\tau_{\bf k}^-$, characterized by the quantum numbers $\nu_{\bf k} = 2$ for ${\bf k} \in \tilde{\mathcal{S}}_2$ and $\nu_{\bf k} = 1(0)$ for ${\bf k} \in \mathcal{S}_{1(0)}$. The resulting projected ground-state energy is given by 
\begin{equation}
E_0(\mu, U, M=0) = \sum_{{\bf k} \in {\mathcal{S}}_1}\tilde{\epsilon
}_{\bf k} + \sum_{{\bf k} \in \tilde{\mathcal{S}}_2}\left(2\tilde{\epsilon
}_{\bf k}+\frac{U}{2}\right).
\label{projectedenergy}
\end{equation}
By comparison with Eq.~\eqref{energyhk}, this can be rewritten as\begin{equation}
E_0(\mu, U, M=0) = \frac{1}{2}E_0'\left(\mu, \frac{U}{5}\right).
\end{equation}
Notice that the above equation implies that the projected ground-state energy of our model, associated with $PH_{\sf n}P$, is simply that of the HK model, up to a global factor of $2$ and a rescaling of $U$ by a factor of $5$. Upon differentiating w.r.t. $\mu$ to obtain $\rho_F$ on both sides, we conclude that the overall factor of $2$ between the two energies simply reflects the fact that the above equation relates the projected ground-state energy of our model for fermion density $\rho_F$ to that of the HK model for fermion density $2\rho_F$. 

In conclusion, aside from an overall rescaling of $\rho_F$ and $U$, the quantum phase diagram of $P H_{\sf n} P$ is identical to that of the HK model at zero temperature. However, the ground states and their degeneracies are vastly different. For further details on the HK model, see Appendix \ref{AppendixHK}.

\section{Single-particle Green's Function and Density of States}
\label{Section4}

\subsection{Retarded Green's Function}
\label{Retarded Green's Function}

In order to study the nature of the single-particle excitations of $H_{\sf n}$ above the ground state, we now turn to the computation of the retarded Green's function of our model. In this section, we focus on the quantity \cite{Economou}
\begin{eqnarray}
\hspace*{-0.5cm}
 G_{\sigma\sigma'}({\bf{x}}, {\bf{x'}}; t, t') = -i \theta(t-t') \left< [ c^{\;}_{{\bf{x}}\sigma}(t), c^{\dagger}_{{\bf{x'}}\sigma'}(t') ]_+\right> ,   
\end{eqnarray}
where $[A,B]_+$ denotes the anticommutator of operators $A$ and $B$. Due to the presence of lattice-translation and time-translation invariance, the above quantity depends only on the differences $\bf{x}-\bf{x'}$ and $t-t'$. It can also be proved that the above quantity is proportional to $\delta_{\sigma \sigma'}$. Hence, the evaluation of the above quantity reduces to computing $\left< [c^{\;}_{\bf{k} \sigma}(t), c^{\dagger}_{\bf{k} \sigma}(0)]_+   \right>$ using the equations of motion method \cite{Zubarev1960}. (Details are explained in Appendix ~\ref{AppendixB}.) The final result for the retarded Green's function in momentum and frequency space is given by 
\begin{eqnarray}
G_{\sigma}({\bf k}, \omega) &=& \frac{A_1}{\omega- \tilde{\epsilon}_{\bf{k}}}+ \frac{A_2}{\omega- \tilde{\epsilon}_{\bf{k}}-\frac{1}{2}U} \nonumber \\&&+ \frac{A_3}{\omega- \tilde{\epsilon}_{\bf{k}}-2U}+ \frac{A_4}{\omega- \tilde{\epsilon}_{\bf{k}}-\frac{5}{2}U} ,
\label{GF-ours}
\end{eqnarray}
where ($G_\uparrow({\bf k}, \omega)=G_\downarrow({\bf k}, \omega)$)
\begin{eqnarray}
A_1 &=& \frac{1}{Z_{\bf{k}}}  \bigg[ 1 + e^{-\beta \tilde{\epsilon}_{\bf{k}}}+ \frac{1}{2}e^{-\beta(2\tilde{\epsilon}_{\bf{k}}+\frac{5U}{2} )}\nonumber + \frac{1}{2} e^{-\beta( 3\tilde{\epsilon}_{\bf{k}}+\frac{5U}{2})}   \bigg] , \nonumber \\
A_2 &=& \frac{1}{Z_{\bf{k}}} \left[\frac{5}{2}e^{-\beta \tilde{\epsilon}_{\bf{k}}} + \frac{5}{2} e^{-\beta(2\tilde{\epsilon}_{\bf{k}}+\frac{U}{2})}  \right] , \nonumber \\
A_3 &=& \frac{1}{Z_{\bf{k}}} \left[\frac{5}{2} e^{-\beta(2\tilde{\epsilon}_{\bf{k}}+\frac{U}{2} )}+ \frac{5}{2} e^{-\beta( 3\tilde{\epsilon}_{\bf{k}}+\frac{5U}{2})}   \right], \nonumber \\
A_4 &=& \frac{1}{Z_{\bf{k}}} \bigg[ \frac{1}{2} e^{-\beta \tilde{\epsilon}_{\bf{k}}} 
+ \frac{1}{2} e^{-\beta(2\tilde{\epsilon}_{\bf{k}} +\frac{5U}{2})} 
+ e^{-\beta( 3\tilde{\epsilon}_{\bf{k}}+\frac{5U}{2})} \nonumber \\
&& \hspace*{1.cm}+ e^{-\beta( 4\tilde{\epsilon}_{\bf{k}}+5U)} \bigg] ,
\end{eqnarray}
with $\beta=1/(k_B T)$ representing the inverse temperature,  measured in units of $2t$, and the partition function corresponding to the subspace ${\cal H}_{\bf k}$ given by
\begin{eqnarray}
Z_{\bf{k}} &=& 1 + 4 e^{-\beta \tilde{\epsilon}_{\bf{k}} } + 5 e^{-\beta (2\tilde{\epsilon}_{\bf{k}}+\frac{U}{2})} + e^{-\beta (2\tilde{\epsilon}_{\bf{k}}+\frac{5U}{2})}  \nonumber \\ 
&& + 4e^{-\beta (3\tilde{\epsilon}_{\bf{k}} + \frac{5U}{2})} + e^{-\beta (4\tilde{\epsilon}_{\bf{k}} + 5U)}.
\end{eqnarray}

The above result is valid for an arbitrary number of spatial dimensions. We note that the poles of the Green's function coincide exactly with the location of the many-body Fermi surfaces, just as in Fermi-liquid theory, even though $H_{\sf n}$ hosts  non-Fermi liquid phases. The form of the Green's function leads us to conclude that our model describes a quadruply-fractionalized Fermi liquid, where each physical electron is decomposed into four emergent quasiparticles  carrying distinct quantum numbers. Those quasi-particles are non-standard fermions, analogous to the holons and doublons in the HK model, with each associated to one of the four poles identified above. 

In addition, we point out that while the quasi-particles of our model might be non-trivial, the quasi-pairs of our model are rather simple. It is easily seen that 
\begin{equation}
\left[H_{\sf n}, \tau_{\bf{k}}^+ \right] = \left( 2\tilde{\epsilon}_{\bf{k}}+ \frac{5U}{2}\right)\tau_{\bf{k}}^+ \ ,
\end{equation}
which establishes the pseudo-spin raising and lowering operators as the required quasi-pairs of our model with infinite lifetimes. This is reminiscent of the phenomenon of $\eta$ pairing in the Hubbard model \cite{Yang1989, HubbardReview2022, Essler1992}.  
\begin{figure*}[htb]
    \centering \includegraphics[width=0.95\textwidth]{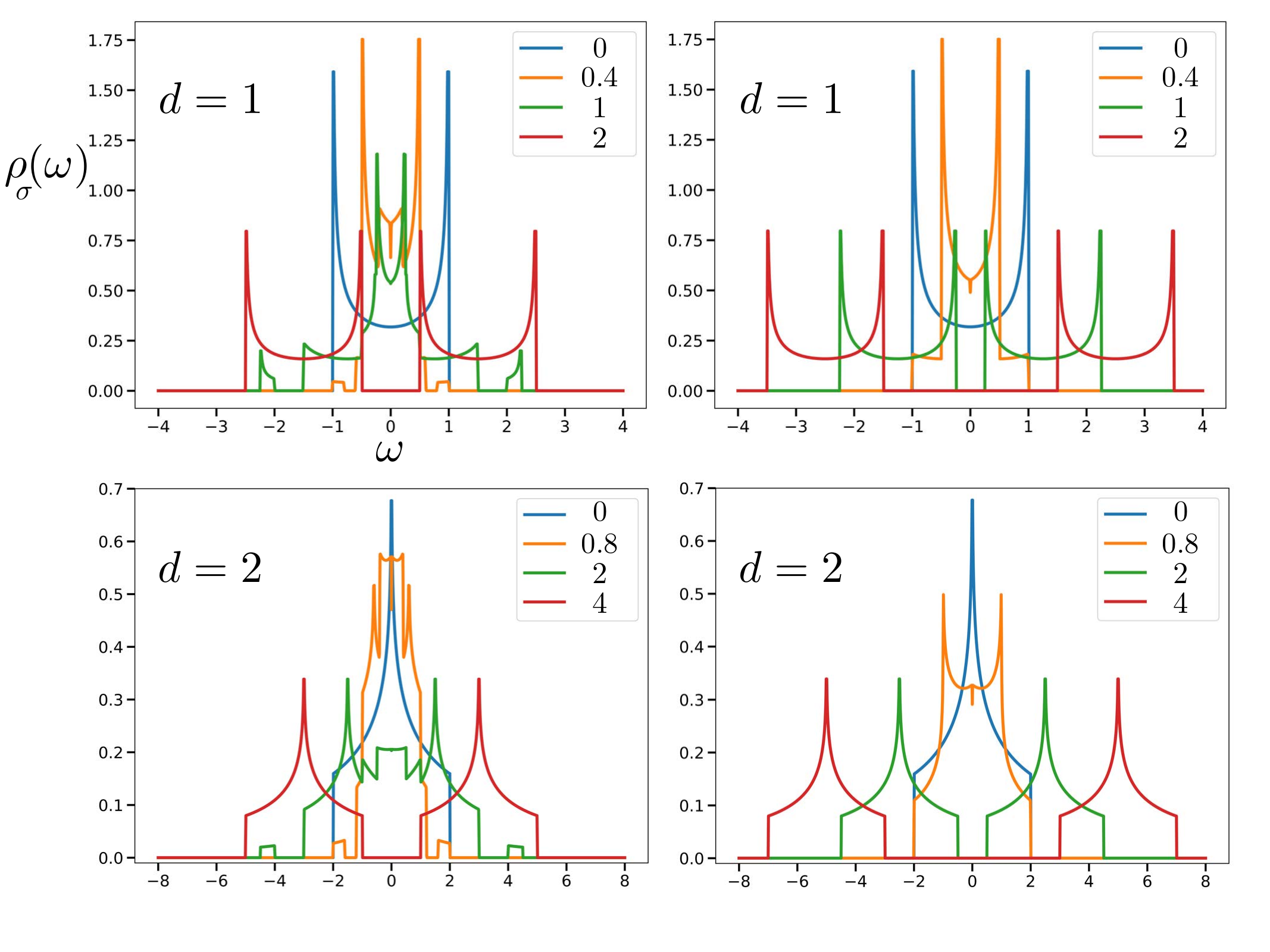}
    \caption{DOS $\rho_{\sigma}(\omega)$, at inverse temperature $\beta=400$,  in $d=1$ (top) and $d=2$ (bottom) dimensions evaluated for selected values of $U$ at half-filling ($\rho_F=1$), for our model (left) and the HK model (right). These plots were calculated in the thermodynamic limit $L \to \infty$ and $\delta = 0$.  For our model,  $U=0.4, 1, 2$ for $d=1$, and $U=0.8, 2, 4$ for $d=2$ correspond to the phases \textquotedblleft(4,3,2,1,0)\textquotedblright, \textquotedblleft(3,2,1)\textquotedblright and \textquotedblleft(2)\textquotedblright,  respectively. For the HK model, $U=0.4, 1, 2$ for $d=1$ and $U=0.8, 2, 4$ for $d=2$ correspond to the phases \textquotedblleft(4,2',0)\textquotedblright, \textquotedblleft(2')\textquotedblright and \textquotedblleft(2')\textquotedblright, respectively. The Fermi liquid DOS at $U=0$ is also shown for comparison. }
    \label{fig:doshalf}
\end{figure*}

\subsection{Single-Particle Density of States}
\label{Single Particle Density of States}

Having calculated the retarded Green's function, we can calculate the single-particle density of states (DOS) 
\begin{equation}
\rho_\sigma(\omega) = -\lim_{\delta \to 0} \frac{2}{\pi V}\sum _{\bf {k}} \text{Im}\, G_{\sigma}({\bf{k}}, \omega+i \delta) ,
\end{equation}
which satifies the sum rule $\int_{-\infty}^{\infty}d\omega \,\rho_\sigma(\omega) =1$. 

The DOS in $d=1$ and $d=2$ at half-filling is shown in Fig. ~\ref{fig:doshalf}. The depletion in $\rho_{\sigma}(\omega=0)$ is clearly visible in both cases as one travels from the metallic phase to the Mott-insulating phase. 

\begin{figure*}[t]
    \centering \includegraphics[width=\textwidth]{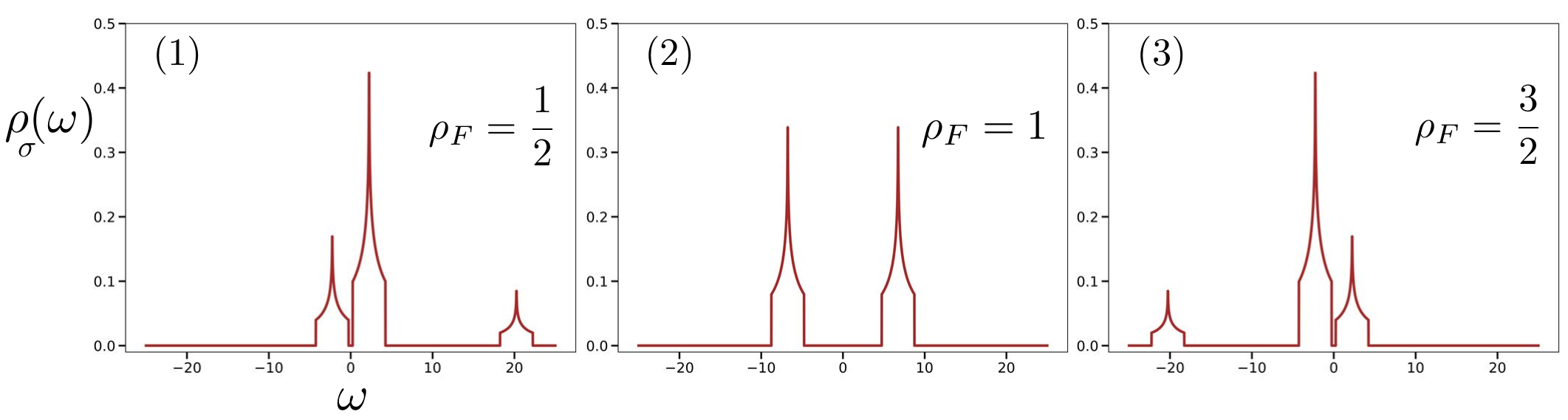}
    \caption{DOS of Mott-insulating  phases, at inverse temperature $\beta=400$,  in $d=2$ dimensions and $U=9$, in the thermodynamic limit $L \to \infty$. The presence of distinct bands separated by Mott-gap(s) is clearly visible.}
    \label{fig:mott_dos}
\end{figure*}

The DOS in the three Mott-insulating phases are shown in Fig.~\ref{fig:mott_dos}. The effect of particle-hole symmetry of $H_{\sf n}$ on the DOS is clearly evident in these plots. Indeed, from the expression of the Green's function in Eq. \eqref{GF-ours}, one can prove that, for fixed $\beta$ and $U$, 
\begin{equation}
\rho_{\sigma}(-\omega)|_{\mu}= \rho_{\sigma}(\omega)|_{\frac{5U}{2}-\mu}  
\end{equation}
in arbitrary spatial dimensions $d$. 

\section{Thermodynamics of $H_{\sf n}$}
\label{Section5}

The grand partition function of our model in arbitrary dimensions is given by $Z(\beta, \mu, U)= {\rm Tr}(e^{-\beta H_{\sf n}}) = \Pi_{\bf{k}} Z_{\bf{k}}$, where the product is over all $\bf k$ with $k_1>0$. Consequently, we can take the grand potential density of our model to be given by 
\begin{eqnarray}\hspace*{-0.7cm}
\Omega(\beta, \mu, U) &=& -\frac{1}{\beta V} \ln Z(\beta, \mu, U) = -\frac{1}{\beta V} \sum_{\bf{k}} \ln Z_{\bf{k}} .
\end{eqnarray}

The fermion density is given by the expression \begin{eqnarray}
    \rho_F = - \frac{\partial}{\partial \mu} \Omega(\beta, \mu, U) ,
\end{eqnarray}
with a maximum value of $2$. 

Using the particle-hole symmetry of $H_{\sf n}$ as written in Eq. \eqref{parthole}, one can prove that 
\begin{equation}
\Omega\left(\beta, \frac{5}{2}U - \mu, U\right)= \Omega(\beta, \mu, U) -\frac{5U}{2}+2\mu.
\label{phnfl}
\end{equation}

Hence, it follows that  
$ \rho_F \rvert_{\mu} + \rho_F \rvert_{\frac{5}{2}U-\mu} = 2$ for arbitrary values of $\beta$ and $U$. Consequently, it is sufficient to study $\rho_F \leq 1$ in order to obtain a complete picture of the thermodynamic phase diagram. 

We next compute some thermodynamic quantities of our model. 

\subsection{Compressibility}
\label{Compressibility}

We can obtain an analytic evaluation for the fermion density at zero temperature, $T=0$,  as given by
\begin{multline}
\rho_F = \frac{1}{V} \sum_{\bf k} \left[ \theta\left( -\tilde{\epsilon}_{\bf{k}}\right) + \theta\left(-\tilde{\epsilon}_{\bf{k}} - \frac{U}{2}\right) \right. \\
\left. + \theta\left(-\tilde{\epsilon}_{\bf{k}} - 2U\right) + \theta\left(-\tilde{\epsilon}_{\bf{k}} - \frac{5U}{2}\right) \right].
\end{multline}
Hence, we can determine the ground-state compressibility  as 
\begin{eqnarray}
\kappa=\frac{d \rho_F}{d \mu} &=& \frac{1}{2}\left[\rho_0(\mu) + \rho_0\left(\mu-\frac{U}{2}\right) \right. \nonumber \\&& \left. + \rho_0(\mu-2U) + \rho_0\left(\mu -\frac{5U}{2}\right) \right],
\label{compressibility}
\end{eqnarray}
where $\rho_0(\mu) = \frac{2}{V}\sum_{\bf k} \delta(- \tilde{\epsilon}_{\bf{k}})$ is the $T=0$ DOS $\rho_{\sigma}(\omega)$ evaluated at $U=\omega=0$. One can interpret Eq.  \eqref{compressibility} as describing a system with four ``Hubbard-like bands" separated by energies $\frac{1}{2}U$ ,$\frac{3}{2}U$ and $\frac{1}{2}U$ respectively. This must be compared to the situation in the HK model, where the relevant expression is given by
\begin{equation}
\kappa_{\sf HK} = \rho_0(\mu) + \rho_0\left(\mu -\frac{5U}{2}\right) ,
\label{compressibilityHK}
\end{equation}
which displays two Hubbard-like bands separated by an energy of $\frac{5}{2}U$.

Equation \eqref{compressibility} also implies that, as in the HK model, the behavior of the compressibility is entirely determined by the behavior of the function $\rho_0(\mu)$. For example, one can prove that as we approach the Mott-insulating phases from the metallic part of the phase diagram, the compressibility behaves as 
\begin{eqnarray}
\kappa &\sim\Big|&\rho_F - \frac{\sf n}{2}\Big|^{1-\frac{2}{d}} \text{ for } \rho_F \rightarrow \frac{\sf n}{2} \text{ and } U>U_{c}(\rho_F) \nonumber ,\\
\kappa &=& 0 \text{ for } \rho_F =\frac{\sf n}{2} \text{ and } U>U_{c}(\rho_F) ,
\end{eqnarray}
where ${\sf n}=1,2,3$ correspond to the Mott-insulating phases \textquotedblleft(1)\textquotedblright, \textquotedblleft(2)\textquotedblright and \textquotedblleft(3)\textquotedblright, respectively, and $U_c(\rho_F)$ is defined in Eq.~\eqref{criticalu}.

\subsection{Energy}
\label{Energy}

\begin{figure*}[t]
    \centering
    \includegraphics[width=0.95\textwidth]{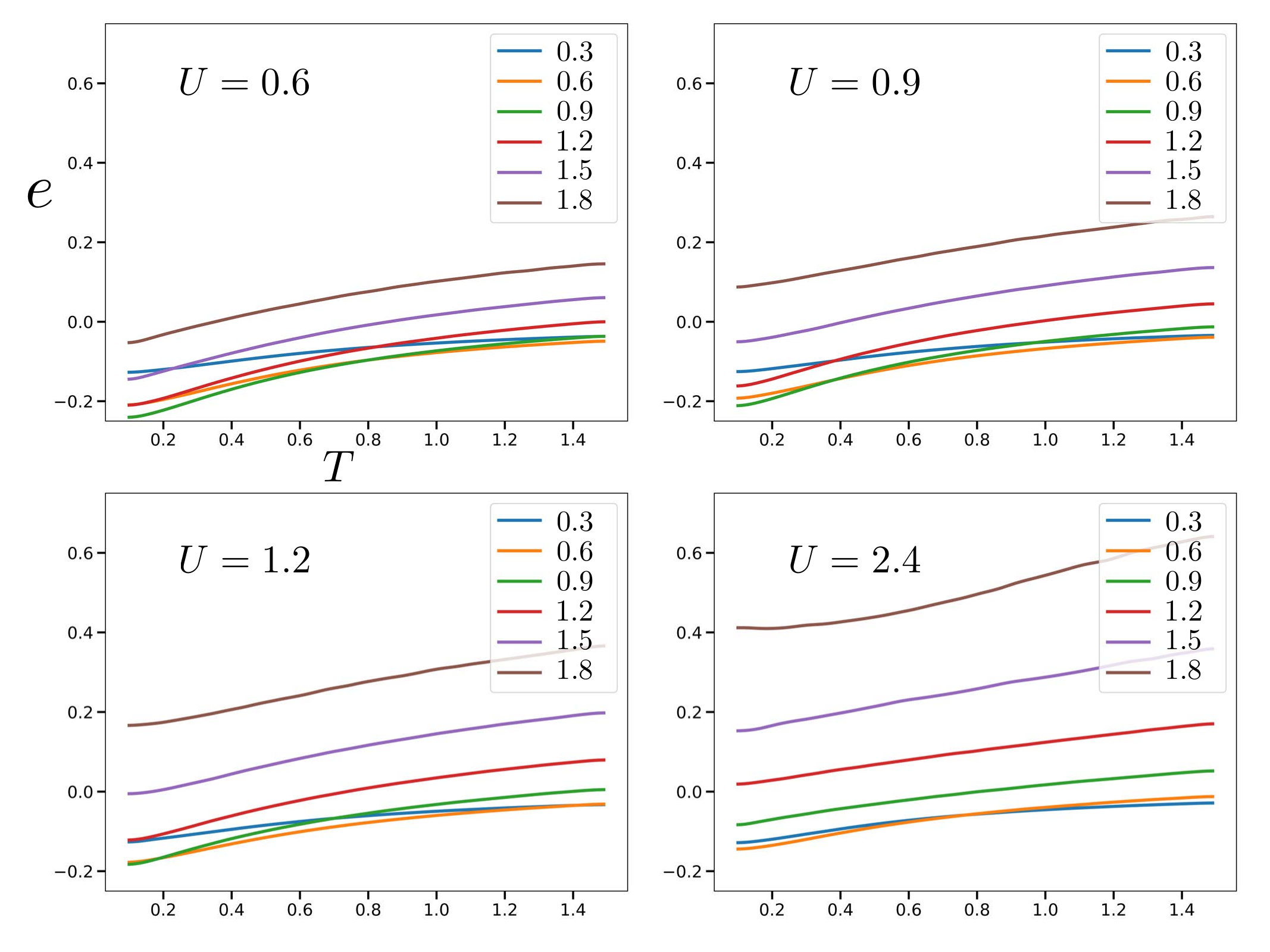}
    \caption{Energy per site $e$ as a function of temperature $T$ for selected values of $U$. Different curves correspond to the values of $2\rho_F$ indicated in the legends.}
    \label{fig:energies}
\end{figure*}

The energy density $e(\beta, \rho_F, U)$ in the canonical ensemble (as a function of $\rho_F$ instead of $\mu$) is given by 
\begin{equation}
e(\beta, \rho_F, U) =  \frac{\partial (\beta \Omega)}{\partial \beta} + \mu \rho_F,
\end{equation}
and is plotted for selected values of $U$ and $\rho_F$ in Fig.~\ref{fig:energies}. Indeed, upon comparing with Eq. \eqref{energydensity} one can easily see that 
\begin{equation}
\lim_{\beta \rightarrow \infty} e(\beta, \rho_F, U) = \epsilon_0(\rho_F, U).
\end{equation}
Additionally, using Eq. \eqref{phnfl} allows us to relate the energy density at $\rho_F$ to that at $2-\rho_F$, keeping $\beta$ and $U$ fixed, as
\begin{equation}
e|_{2-\rho_F} = e|_{\rho_F} -\frac{5}{2}U(\rho_F-1),
\label{enersymm}
\end{equation}
which is a consequence of Eq \eqref{phnfl}, reflecting the particle-hole symmetry of our system. 

Interestingly, at large values of $U$, in an arbitrary number of spatial dimensions $d$, there are additional ``emergent particle-hole symmetries" as $T\rightarrow0$. This is a familiar feature also seen in the HK model. Note that if $U>4d$, the only metallic phases that appear are \textquotedblleft(1,0)\textquotedblright, \textquotedblleft(2,1)\textquotedblright, \textquotedblleft(3,2)\textquotedblright and \textquotedblleft(4,3)\textquotedblright. In each of these phases, there exists only one many-body Fermi surface in the half-BZ, whose location is uniquely fixed by the fermion number density $\rho_F$, independently of $U$ ($U>4d$). This implies that, in these phases, the many-body ground state given in Eq. \eqref{gsofhn} is entirely determined by $\rho_F$ and does not depend on $U$, which in turn leads to a ground-state energy that is linear in $U$. From the particle–hole symmetry of the non-interacting band dispersion, an emergent particle–hole symmetry follows, which we summarize in Table \ref{phtable}.
\begin{figure*}[t]
    \centering
    \includegraphics[width=0.95\textwidth]{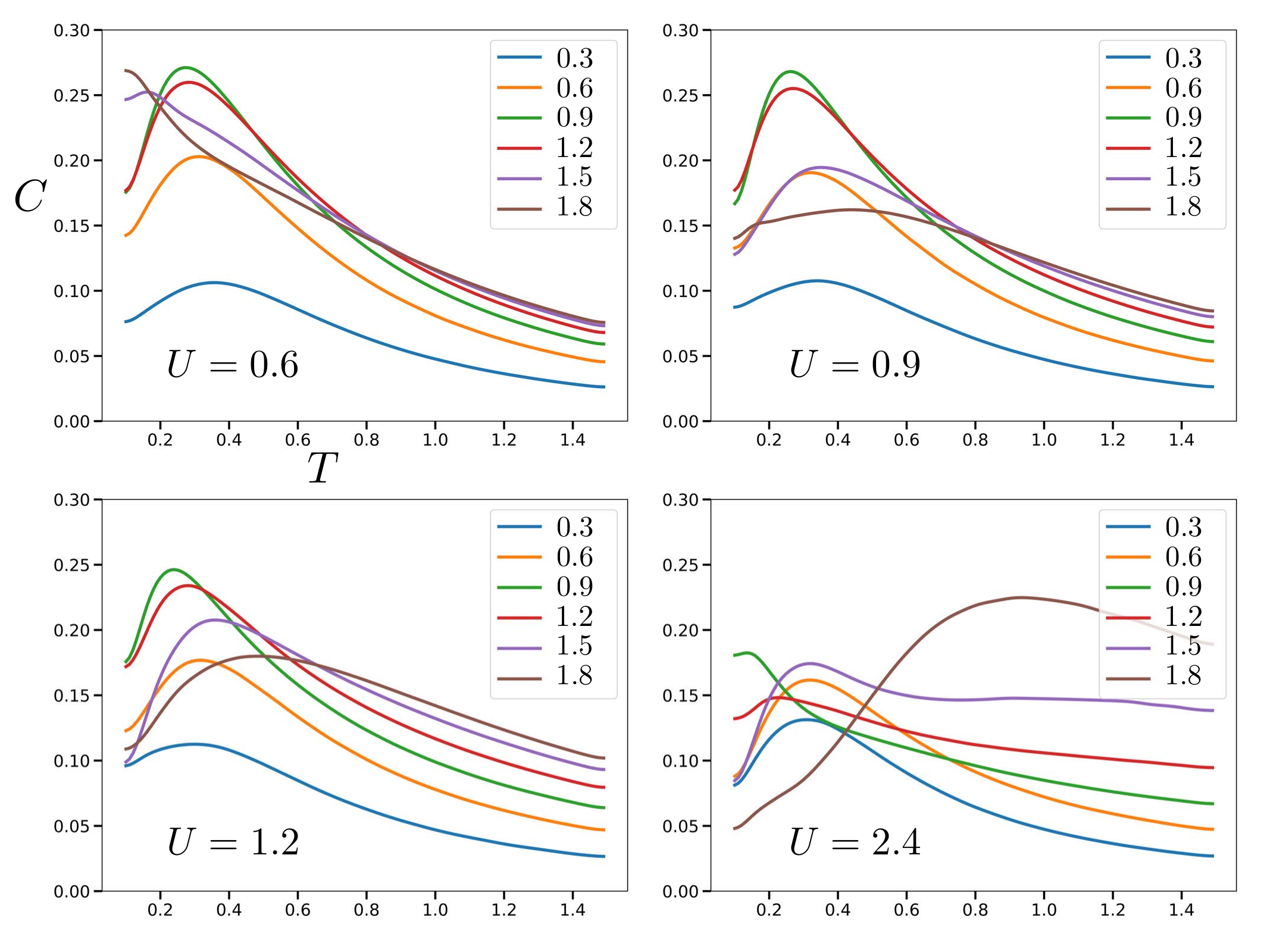}
    \caption{Specific heat $C$ as a function of temperature $T$ for selected values of $U$.  Different curves correspond to the values of $2\rho_F$ indicated in the legends. }
    \label{fig:spheats}
\end{figure*}

\begin{table}[htb]
\[
\begin{array}{|c|c|c|c|c|}
\hline
 & \text{Phase} & \rho_F' & \epsilon_0(\rho_F', U) - \epsilon_0(\rho_F, U) & \text{Range of } \rho_F\\
\hline
1 & \text{ \textquotedblleft (1,0)\textquotedblright} & \frac{1}{2}-\rho_F & 0 & 0<\rho_F < \frac{1}{2}\\
2 & \text{ \textquotedblleft (2,1)\textquotedblright}& \frac{3}{2}-\rho_F & U\left( \frac{3}{4}-\rho_F  \right) & \frac{1}{2}< \rho_F< 1\\
3 & \text{ \textquotedblleft (3,2)\textquotedblright} & \frac{5}{2}-\rho_F&4U\left( \frac{5}{4}-\rho_F  \right)& 1<\rho_F<\frac{3}{2}\\
4 & \text{ \textquotedblleft (4,3)\textquotedblright} & \frac{7}{2}-\rho_F  & 5U\left( \frac{7}{4}-\rho_F  \right)& \frac{3}{2}<\rho_F<2\\
\hline
\end{array}
\]
\caption{Emergent particle-hole symmetries, at $T=0$, relating the ground-state energy $\epsilon_0(\rho_F, U)$ at different fillings $\rho_F$ and $\rho_F'$ for $U>4d$. } 
\label{phtable}
\end{table}

\subsection{Specific Heat}
\label{Specific Heat}

The specific heat of our model can be calculated by the expression
\begin{equation}
C(\beta, \rho_F, U) = \frac{\partial e}{\partial T}\Bigg|_{\rho_F}.
\end{equation}
This is plotted for selected values of $U$ and $\rho_F$ in Fig.~\ref{fig:spheats}. Differentiating Eq.  \eqref{enersymm} with respect to temperature yields the particle-hole symmetric result
\begin{equation}
C|_{\rho_F} = C|_{2-\rho_F}.
\end{equation}
Further numerical analysis suggests that the specific heat scales linearly, that is, $C \propto T $, at low temperatures across all regions of the phase diagram featuring non-Fermi liquid phases, consistent with gapless excitations characteristic of metallic behavior. In contrast, within the Mott-insulating phases, $ C $ displays exponential suppression at low temperatures, in agreement with the presence of a finite charge gap, $\Delta \mu \neq 0$.
\begin{figure*}[htb]
    \centering
    \includegraphics[width=0.95\textwidth]{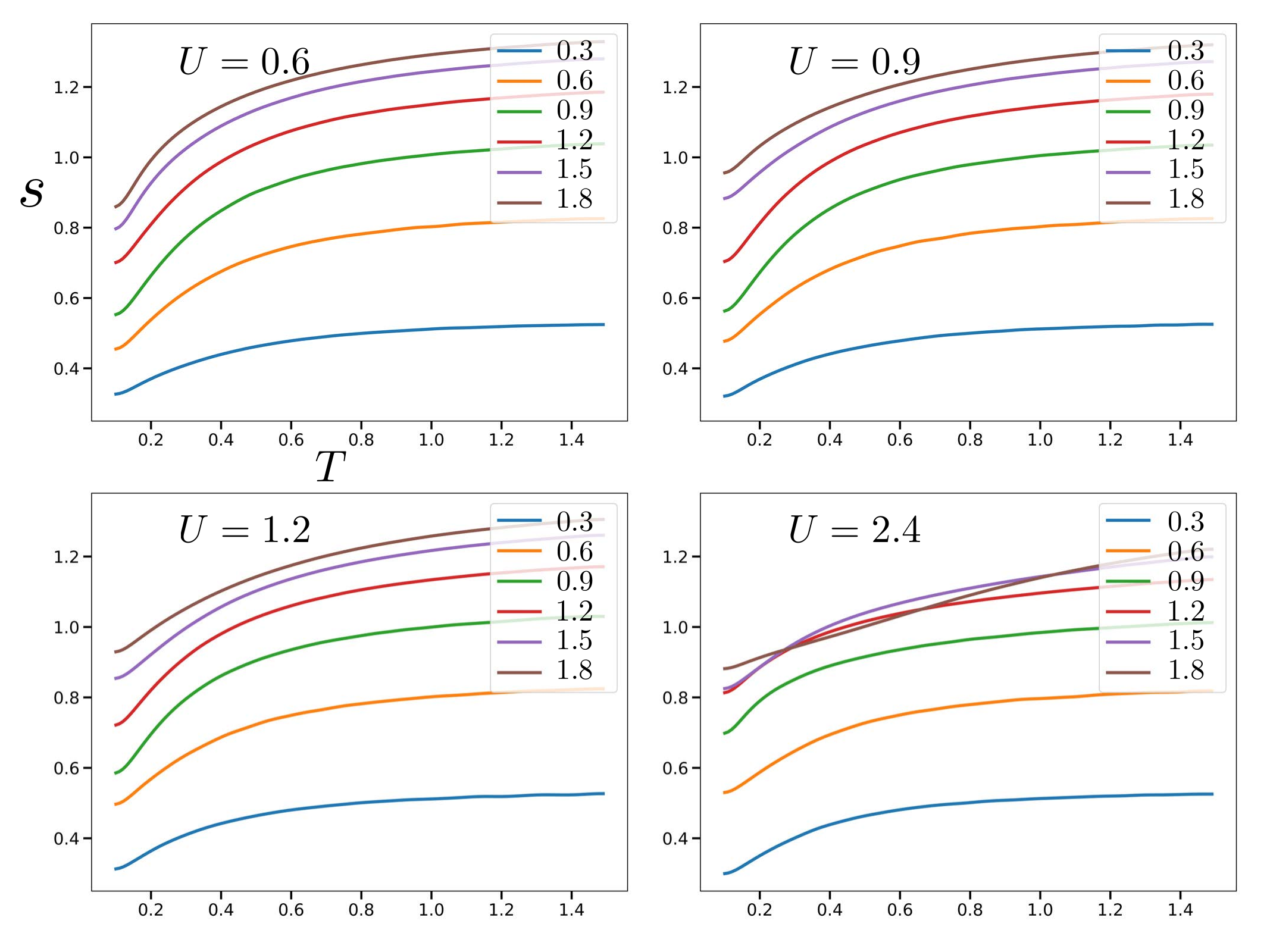}
    \caption{Entropy per site $s$ as a function of temperature $T$ for selected values of $U$. Different curves correspond to the values of $2\rho_F$ indicated in the legends.}
    \label{fig:entropies}
\end{figure*}

\subsection{Entropy}
\label{Entropy}

The entropy density is given by 
\begin{equation}
s(\beta, \rho_F, U) = \beta (e(\beta, \rho_F, U)-\Omega(\beta, \rho_F, U) - \mu \rho_F),
\end{equation}
and is shown for selected values of $U$ and $\rho_F$ in Fig.~\ref{fig:entropies}. The particle-hole symmetry of $H_{\sf n}$ manifests itself in  $s(\beta, \rho_F, U)$ via the symmetric expression
\begin{equation}
s|_{\rho_F}= s|_{2-\rho_F} .
\end{equation}
The low temperature limit of the entropy density is dominated by the macroscopic degeneracy of the ground state of $H_{\sf n}$, Eq.  \eqref{gsofhn}. Precisely,
\begin{equation}
\lim_{T \rightarrow 0} s = \frac{1}{V}\{|\mathcal{S}_1|\text{ ln }4 +|\mathcal{S}_2|\text{ ln }5+|\mathcal{S}_3|\text{ ln }4\} ,
\end{equation}
where $\mathcal{S}_i$ are the sets $\mathcal{S}_i$ used to describe the ground state, as discussed in Eq. \eqref{fermisurfaces}. It can be easily seen that for $U>0$, $\rho_F \neq 0,2$, at least one among $|\mathcal{S}_1|$, $|\mathcal{S}_2|$ and $|\mathcal{S}_3|$ scales with $V$ with a positive coefficient, thus establishing that our model, like the HK model, has a nonzero value of the entropy density, at zero temperature, for \textit{arbitrary} values of $U>0$ and $\rho_F \neq 0,2$.

\section{Cooper Pair Instability}
\label{Section6}

We now examine the stability of the various non-Fermi liquid ground states of $H_{\sf n}$ in the presence of the pairing interaction. Specifically, we explore whether two electrons can form a bound state at zero temperature under an arbitrarily weak attractive interaction ($G>0$) — an idea that may offer key insights into the emergence of different types of superconducting order parameters. We want to study the potential instability as a function of $\mu, U$ and $G$.
In order to do so, we focus on the two-body Hamiltonian $H$ and consider the competition between the $M=0$ and $M=1$ sectors of the Hilbert space for the ground state in the thermodynamic limit $L_1=L_2 \cdots=L_d=L\rightarrow \infty$. 

Having already calculated the energy of the states belonging to the sector $M=0$ in Section ~\ref{HKemergence}, let us turn our attention to the $M=1$ sector. In this sector, the Cooper pair equation \cite{Cooper-pair} for $x_1=x$, which incorporates the repulsive interaction $U$ through the choice of $\bf k$ vectors, takes the form of the Richardson equation for a single pairon 
\begin{equation}
\sum_{{\bf k} \in \mathcal{S}_0}\frac{1}{\tilde{\epsilon
}_{\bf k}-x} + \frac{1}{2}\sum_{{\bf k} \in \mathcal{S}_1}\frac{1}{\tilde{\epsilon
}_{\bf k}-x}=\frac{1}{G},
\label{singlepaironrg}
\end{equation}
which allows us to write the total energy functional as 
\begin{equation}
E\left(\mu, U,G,M=1\right) = E(\mu, U, M=0)  +2x+\frac{5U}{2} , 
\end{equation}
where $E(\mu, U, M=0)$ is defined in Eq. \eqref{projectedenergy}, and $\mathcal{S}_0$ and $\mathcal{S}_1$ are the sets defined in Eq \eqref{k-sectors}. One can prove that the smallest root,  $x=\tilde{x}$, of the Cooper pair equation varies continuously from $-\infty$ to $\tilde{\epsilon}_2$ as $G$ decreases from $\infty$ to $0$, with $\mu$ and $U$ held fixed. For further details, see Appendix \ref{Onepairapp}.


In order to investigate the possibility of multiple superconducting phases in our model, let us now investigate the behavior of the \textit{binding energy}, $E_b$, defined as 
\begin{eqnarray}
E_b &=& E(\mu, U,G, M=1)-2\tilde{\epsilon}_2 -\frac{5U}{2} \nonumber \\
&& - E(\mu, U, M=0) ,\nonumber \\
&=& 2\tilde{x} -2\tilde{\epsilon}_2 .
\end{eqnarray}

To this end, let us define $g\equiv \frac{1}{2}GV$ and set $d=1$. First, we focus on the case of $\rho_F=\frac{1}{2}$ i.e., $\mu=U/4$ for $U<4$. This corresponds to the phase \textquotedblleft(2,1,0)\textquotedblright. In this case,  $\tilde{x}$ satisfies the equation
\begin{equation}
\text{tanh}^{-1} \sqrt{\frac{1-\left(\frac{U}{4}+\tilde{x}   \right)^2}{1-\left( \frac{U}{4}  \right)^2}} = \frac{\pi}{g} \sqrt{ 1-\left(\frac{U}{4}+\tilde{x}   \right)^2 },
\label{210sc}
\end{equation}
which can be simplified for infinitesimal $g$ to give us a binding energy of the form
\begin{equation}
E_b \sim \frac{U}{2} -2\, \sqrt{1-u + 4u\text{ exp} \left( -\frac{2 \pi}{g}  \sqrt{u} \right)} , 
\end{equation}
where $u= 1-\left(\frac{U}{4}\right)^2$. 

Next, let us focus on the phase \textquotedblleft(2,1)\textquotedblright. Specifically, we set $\rho_F=\frac{3}{4}$,  i.e., $\mu=U/2$ for $U>2$. In this case, $\tilde{x}$ satisfies the equation
\begin{equation}
\text{tanh}^{-1} \sqrt{\frac{1+\frac{U}{2}+\tilde{x}}{1-\frac{U}{2}-\tilde{x}}} = \frac{\pi}{g}\sqrt{1-\left(\frac{U}{2}+\tilde{x}   \right)^2},
\label{21sc}
\end{equation}
which can be simplified for infinitesimal $g$ to give us a binding energy of the form
\begin{equation}
E_b \sim  -4 \text{ exp}\left(-\frac{2\pi}{g}  \right) .
\end{equation}
It must be noted that in order to find $\tilde{x}$ for finite $g$, Eqs. \eqref{210sc} and \eqref{21sc} must be extended to include the cases in which the arguments of the square roots are negative, in which case the inverse hyperbolic tangent can be rewritten as an inverse tangent for the price of a factor of $i$. 

Next, we focus on the phase \textquotedblleft(1,0)\textquotedblright. Specifically, we set $\rho_F=\frac{1}{4}$,  i.e., $\mu=0$ for $U>2$. In this case, $\tilde{x}$ is the smallest root of the equation
\begin{equation}
\frac{\pi}{2}+ \text{tan}^{-1}\sqrt{\frac{\tilde{x}+1}{\tilde{x}-1}} = \frac{\pi}{g}\sqrt{\tilde{x}^2-1} ,
\end{equation}
which can be simplified for infinitesimal $g$ to yield a binding energy of the form
\begin{equation}
E_b \sim 2\left( 1-\sqrt{1+\frac{g^2}{4}} \right).
\end{equation}

In the Mott-insulating phase \textquotedblleft(1)\textquotedblright, i.e., $\rho_F=\frac{1}{2}, U>4$, and $\tilde{x}$ is given by 
\begin{equation}
\tilde{x} = -\frac{U}{4}-\sqrt{1+\frac{g^2}{4}},
\end{equation}
which leads us to the exact binding energy
\begin{equation}
E_b = 2 \left(1- \sqrt{1+\frac{g^2}{4}}  \right).
\end{equation}





Notice that the binding energy corresponding to the phase \textquotedblleft(2,1,0)\textquotedblright is different from the standard BCS result. In addition, the binding energies corresponding to the two non-Fermi liquid phases \textquotedblleft(2,1,0)\textquotedblright and \textquotedblleft(2,1)\textquotedblright display essential singularities with different coefficients in the exponential,  indicating the potential existence of \textit{two} different superconducting instabilities. On the contrary, the binding energies corresponding to the Mott insulating phase \textquotedblleft(1)\textquotedblright and the non-Fermi liquid phase \textquotedblleft(1,0)\textquotedblright have no essential singularities, suggesting the absence of a superconducting instability in these two phases.  

Finally, we note that in order to describe the ground state for $\rho_F> 1$, we can take advantage of the particle-hole symmetry of the Hamiltonian. One can easily see that the particle-hole operator $\mathcal{K}_{ph}$ maps the pairon vacua $|\lambda\rangle$ to the anti-pairon vacua, $|\lambda_{ph}\rangle = \mathcal{K}_{ph}|\lambda\rangle $, (which are annihilated by all the $\tau_{\bf k}^+$) and the pairon creation operator $B_{\alpha}^{\dagger}$ to the pairon annihilation operator $B_{\alpha}$, upto a redefinition of $x_{\alpha}$. Consequently, it is sufficient to study $\rho_F\leq 1$ to obtain a complete picture of the binding energy for the entire phase diagram.

\section{Concluding Remarks}
\label{Section7}

In this paper, we introduce a model of interacting fermions that satisfies three remarkable properties: 
\begin{itemize}
    \item The model Hamiltonian is integrable in an arbitrary number of spatial dimensions.
    \item The model simultaneously hosts non-Fermi liquid, Mott insulating, and superconducting phases.
    \item The fundamental excitations of the model are nonstandard \textit{fractionalized} quasiparticles.
\end{itemize}

Our model Hamiltonian consists of a single-particle band contribution, along with repulsive and pairing interaction terms.  We demonstrate that the exact eigenstates and eigenvalues of our Hamiltonian can be obtained with algebraic complexity for arbitrary spatial dimensions and arbitrary interaction strengths, thereby establishing its exact solvability. More specifically, it belongs to the class of Richardson-Gaudin integrable models. The spectrum displays a macroscopic degeneracy as a result of a {\it local} (in momentum space) conservation law. We also establish that the model possesses particle-hole symmetry and explicitly construct the {\it unitary} operator that implements it. Although this work focuses on s-wave pairing, the construction can be readily generalized to $p+ip$ and $so(5)$ versions as well. 

In the absence of pairing terms, the metallic ground states of our model are characterized by a set of \textit{four} many-body Fermi surfaces whose positions can be calculated as a function of the fermion density (or equivalently, the chemical potential) and the repulsion strength. The existence of multiple many-body Fermi surfaces challenges Luttinger’s theorem and signals the breakdown of Landau Fermi liquid theory. Consequently, the quantum metallic phases can be characterized by the structure of their underlying many-body Fermi surfaces. We find a total of \textit{ten} non-Fermi liquid phases and \textit{three} Mott-insulating phases. In two spatial dimensions, the non-Fermi liquid phases also feature lines of many-body Lifshitz (third order) transitions across the phase diagram. These lines correspond to topological changes in the many-body Fermi surfaces. It is particularly interesting to compare this structure to that of the Hatsugai-Kohmoto model (in the absence of pairing terms), which —despite sharing the same eigenbasis— exhibits macroscopically degenerate ground states characterized by {\it two} many-body Fermi surfaces. This degeneracy gives rise to {\it three} distinct non-Fermi liquid phases and a {\it single} Mott insulating phase. Additionally, we find another rather unsurprising result:  When projected onto the subspace consisting of no {\it pairon} excitations, the ground-state energy of our model coincides with that of the Hatsugai-Kohmoto (in the full Hilbert space) for rescaled values of the parameters. 

Indeed, the similarities of our model in the absence of pairing terms to the Hatsugai-Kohmoto extend far beyond the presence of a common eigenbasis. This is captured by our investigation into the nature of excitations above the ground state, which can be characterized by the poles of the retarded single-particle Green’s function, which we compute {\it exactly} using the equations of motion method. We find that the four poles of the Green’s function coincide precisely with the positions of the four many-body Fermi surfaces, indicating that the model hosts nonstandard fractionalized quasiparticles, similar to the holons and doublons of the Hatsugai-Kohmoto model, with infinite lifetimes.

To investigate possible superconducting instabilities of the {\it normal} ground states of our model, we perform a single-pairon analysis, i.e., we address the Cooper-pair problem \cite{Cooper-pair}, taking into account the repulsive interaction. Notably, the binding energy exhibits qualitatively different behaviors across the quantum phase diagram —essential singularities emerge in some regions, whereas others display regular behavior. These observations suggest the presence of distinct superconducting and non-superconducting phases in the model. We defer a detailed exploration into the finite temperature and superconducting phases for future work. 

Our findings reveal an interesting exactly-solvable model  in which Mott physics, non-Fermi liquid behavior, and superconductivity compete and may coexist, offering a unified framework for investigating these phenomena in strongly correlated matter.

\section{Acknowledgements}

We are indebted to Philip Phillips for insightful conversations regarding the role of an emergent $\mathbb{Z}_2$ symmetry as a signature of Mott physics.  G.O. gratefully acknowledges support from the Institute for
Advanced Study. We further acknowledge financial support from Grant PID2022-136992NB-I00 funded by MCIN/AEI/10.13039/501100011033.

\appendix

\section{Extension of the integrable model to a $p_x + ip_y$ topological superconductor}
\label{AppendixE}

As mentioned in Section \ref{Quantum Integrability}, our model can be extended to the hyperbolic
family of RG models \cite{Class2001, Colloquium2004, Ortiz2005} for which one of the most important realizations is the $p+ip$ model of spinless fermions in $d=1$
\cite{Ortiz2014,ORTIZ2016357, Gritsev2019} and the $p_x+ip_y$ model in $d=2$ \cite{Ibanez2009, Rombouts2010, Foster2013, Hazzard2017}.

Following \cite{Rombouts2010}, the $p_{x}+ip_{y}$ model is written in terms of spinless fermions \cite{Ortiz2005}. Additionally, one must include a phase factor in the
definition of pseudo-spin operators, $\tau_{\mathbf{k}}^{+}=\frac{k_{x}%
+ik_{y}}{\left\vert \mathbf{k}\right\vert }c_{\mathbf{k}}^{\dagger
}c_{-\mathbf{k}}^{\dagger} = (\tau^-_{\bf k})^{\dagger}$, 
$\tau_{\mathbf{k}}^{z}=\frac{1}{2}\left(  c_{\mathbf{k}}^{\dagger
}c_{\mathbf{k}}+c_{-\mathbf{k}}^{\dagger}c_{-\mathbf{k}}-1\right) = \frac{1}{2}(\hat{N}_{\bf k} -1)$. It can be readily seen that the addition of this phase does not modify the $su(2)$ algebra.

The $p_{x}+ip_{y}$ Hamiltonian can be written as $H(U=0)= \sum_{\mathbf{k}}\epsilon_{\mathbf{k}}\tau_{\mathbf{k}}^{z} + H_{\sf p}$, where 
\begin{equation}
H_{\sf p}= -G\sum_{\mathbf{k,k}^{\prime}}\sqrt{\epsilon_{\mathbf{k}}\epsilon
_{\mathbf{k}^{\prime}}}\tau_{\mathbf{k}}^{+}\tau_{\mathbf{k}^{\prime}}^{-}.
\label{HP}
\end{equation}

The complete set of eigenstates of the pairing Hamiltonian $H(U=0)$ is given by the Richardson-Gaudin ansatz. While the $s$-wave pairing term studied in detail in our work falls into the rational class of the Richardson-Gaudin models, the $p_x + ip_y$ term above falls into the hyperbolic class. Subsequently, the hyperbolic ansatz for the eigenstates generalizing Eq. \eqref{Ansatz} is given by
\begin{equation}
\left\vert \Psi\right\rangle =
{\displaystyle\prod\limits_{\alpha=1}^{M}}
B_{\alpha}^{\dagger}\left\vert \lambda\right\rangle ,\text{ }B_{\alpha}^{\dagger}=\sum_{\mathbf{k}}\frac{\sqrt{\epsilon_{\mathbf{k}}}}{\epsilon_{\mathbf{k}}
-x_{\alpha}}\tau_{\mathbf{k}}^{+} ,
\label{Hansatz}
\end{equation}
where $|\lambda\rangle$ is the pairon vacuum satisfying the properties $\tau_{\bf k}^{-} |\lambda\rangle = 0$ and $\hat{N}_{\bf k} |\lambda \rangle = \nu_{\bf k} |\lambda \rangle$ for all $\bf k$. In order to ensure that the ansatz above satisfies the eigenvalue equation for $H(U=0)$, the $M$ pairon energies $\left\{  x_{\alpha}\right\}  $  must satisfy the set of $M$ nonlinear Richardson equations
\begin{equation}
\frac{1}{2}\sum_{\mathbf{k}}\frac{1-\nu_{\mathbf{k}}}{\epsilon_{\mathbf{k}%
}-x_{\alpha}}-\sum_{\beta\left(  \neq\alpha\right)  }\frac{1}{x_{\beta
}-x_{\alpha}}-\frac{Q}{x_{\alpha}}=0 ,
\label{Reqn}
\end{equation}
where
\begin{equation}
Q=\frac{1}{2G}-\frac{1}{2}\sum_{\bf k} (1-\nu_{\bf k})+M-1 .
\end{equation}

The $p_x+ip_y$ superfluid  Hamiltonian with repulsion is
\begin{equation}
H\equiv H(U=0) + H_{\sf U}  ,
\end{equation}
where $H_{\sf U}$ is the repulsive interaction given by 
\begin{equation}
H_{\sf U} = U\sum_{\bf k} \left[\tau^+_{\bf k}\tau^-_{\bf k} + \frac{1}{4}\hat{N}_{\bf k}(\hat{N}_{\bf k}-1)\right].     
\end{equation}
Similar to the rational case, we have
\begin{equation}
\left[  H(U=0),H_{\sf U}\right]  =0\text{, \ }\left[ H_{\sf U},B_{\alpha}^{\dagger} \right]
=\frac{3}{2} U B_{\alpha}^{\dagger}\text{.}
\end{equation}
Consequently, $H_{\sf U}$ is a constant of motion whose value is given by
\begin{equation}
H_{\sf U}\left\vert \Psi\right\rangle = \left[  \frac{3}{2}UM+\frac{1}{4}U\sum_{\bf k}\nu_{\bf k}\left(
\nu_{\bf k}-1\right)  \right]  \left\vert \lambda\right\rangle .
\end{equation}

Consequently, the  eigenvalue associated to the eigenstate  \eqref{Hansatz} is
\begin{eqnarray}
E(N, U, G, M,\{\nu_{\bf k}\})&=&\frac{1}{2}\sum_{\mathbf{k}}\epsilon_{\mathbf{k}}(\nu_{\mathbf{k}}-1)
+\sum_{\alpha}x_{\alpha} \nonumber \\ && \hspace*{-0.7cm}+\frac{3}{2}UM +\frac{U}{4}\sum_{\bf k}\nu_{\bf k}\left(  \nu_{\bf k}-1\right) ,
\end{eqnarray}
where $N=2M+\sum_{\bf k} \nu_{\bf k}$.

Despite the similarities in the exact eigenstates of the rational and hyperbolic models, the differing properties of their superfluid phases arise from differences in the Richardson equations, Eqs.~\eqref{Equations} and \eqref{Reqn}, for the $p_{x}+ip_{y}$ case. The  $p_{x}+ip_{y}$ ground state solution describes a topological superfluid for weak values of the pairing strength $G$ (weak pairing). At a critical value $G_{c}$, the system undergoes a third order
quantum phase transition from a topological superfluid (weak pairing) to a 
trivial superfluid state (strong pairing) \cite{Rombouts2010}. It is important to emphasize that all previous studies were restricted to $\nu = 0$ states. However, the inclusion of finite seniority states, as required by our model, could give rise to exotic superfluid phases such as the Sarma or FFLO states \cite{Dukelsky2006}.

\section{Comparison with the Hatsugai-Kohmoto Model}
\label{AppendixHK}

Our model exhibits remarkable similarities with another integrable model of a non-Fermi liquid, the HK model. Indeed, the HK repulsive two-body term constitutes a component of the overall repulsive interaction in our model. In this section, we compare and contrast the two models. 

\subsection{Model Hamiltonian and Eigenstates}

\begin{figure}[t]
    \centering
    \includegraphics[width=0.45\textwidth]{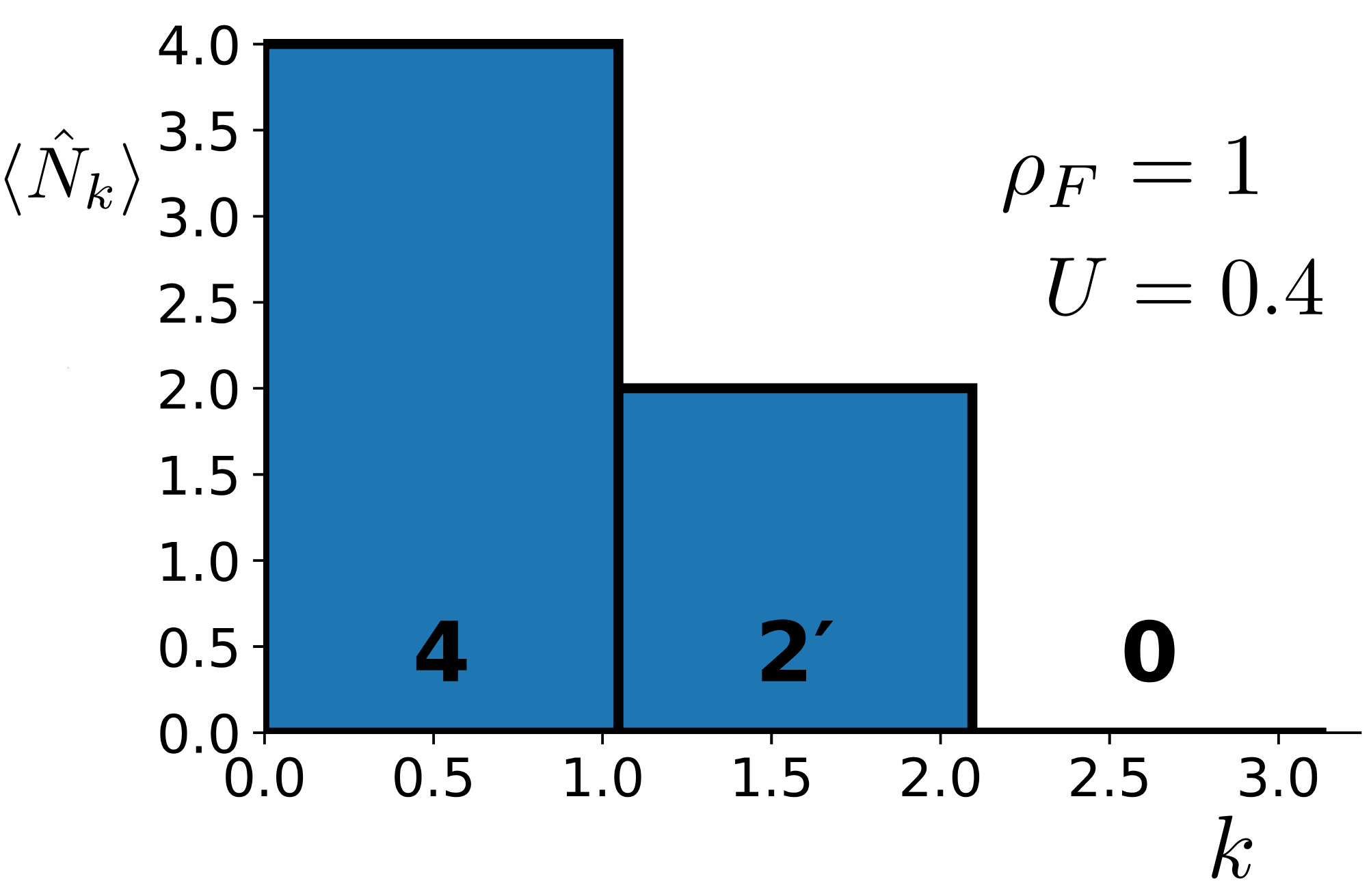}
    \caption{Occupation numbers for the ground state of $H_{\sf HK}$ in phase \textquotedblleft(4,2',0)'' in $d=1$. }
    \label{fig:nkhkd}
\end{figure}

To facilitate comparison with a conventional model for a non-Fermi liquid, we rewrite the HK model in momentum space as follows
\begin{equation}
H_{\sf HK} = \sum_{\bf{k}} \tilde{\epsilon}_{\bf{k}} \hat{N}_{\bf{k}} + \frac{5}{2}U \sum_{\bf{k}} \left( n_{\bf{k} \uparrow} n_{\bf{k} \downarrow} +  n_{-\bf{k} \uparrow} n_{-\bf{k} \downarrow}  \right) ,
\end{equation}
where the summation extends over all $\bf{k}$ vectors with ${k}_1>0$, and $\hat{N}_{\bf{k}}$ is the same number operator in Eq. \eqref{nonfermi}. Consequently, the free-fermion term is exactly the same as in our model, and the factor of $\frac{5}{2}$ is introduced into the second term to ensure that the energy of the fully filled state ($\rho_F=2$) is the same for $H_{\sf n}$ and $H_{\sf HK}$ for an arbitrary value of $U$.
Written in this way, it can be verified that any eigenstate of our model is also an eigenstate of the HK model. Written below in Table \ref{table5} are the energies for the various eigenstates of $H_{\sf HK}$.
\begin{table}[htb]
\[
\begin{array}{|c|c|c|c|c|}
\hline
 & M_{\bf k}, s^z_{\bf k}, T_{\bf k}, \nu_{\bf k} & \text{State}& \text{$\tilde E^{\sf HK}_{\bf k}(\mu,U)$}\\
\hline
1 & 0, 0, 0, 0 & |0\rangle & 0 \\
2 & 0, \frac{1}{2}, {\bf k}, 1 & c_{{\bf k}\uparrow}^\dagger  |0\rangle & \tilde\epsilon_{\bf{k}}\\
3 & 0, -\frac{1}{2}, {\bf k}, 1 & c_{\bf k  \downarrow}^\dagger |0\rangle &\tilde\epsilon_{\bf{k}}\\
4 & 0, \frac{1}{2}, -{\bf k}, 1 & c_{-{\bf k}\uparrow}^\dagger  |0\rangle & \tilde\epsilon_{\bf{k}}\\
5 & 0, -\frac{1}{2}, -{\bf k}, 1 & c_{-{\bf k}\downarrow}^\dagger  |0\rangle & \tilde\epsilon_{\bf{k}} \\
6 & 0, 0, 2{\bf k}, 2 & c_{\bf k \uparrow}^\dagger  c_{\bf k  \downarrow}^\dagger |0\rangle & 2\tilde\epsilon_{\bf{k}} + \frac{5U}{2}\\
7 & 0, 1, 0, 2 & c_{\bf k \uparrow }^\dagger c_{-{\bf k}\uparrow}^\dagger  |0\rangle & 2\tilde\epsilon_{\bf{k}}\\
8 & 0, -1, 0, 2 & c_{\bf k  \downarrow}^\dagger c_{-{\bf k}\downarrow}^\dagger  |0\rangle &2\tilde\epsilon_{\bf{k}} \\
9 & 0, 0, -2{\bf{k}}, 2 & c_{-{\bf k}\downarrow}^\dagger  c_{-{\bf k}\uparrow}^\dagger  |0\rangle & 2\tilde\epsilon_{\bf{k}} + \frac{5U}{2}\\
10 & 0, 0, 0, 2 & \frac{1}{\sqrt{2}}(c_{\bf k \uparrow}^\dagger  c_{-{\bf k}\downarrow}^\dagger  + c_{\bf k \downarrow}^\dagger  c_{-{\bf k} \uparrow}^\dagger ) |0\rangle & 2\tilde\epsilon_{\bf{k}}\\
11 & 1, 0, 0, 0 & \frac{1}{\sqrt{2}}\tau_{\bf k}^+ |0\rangle & 2\tilde\epsilon_{\bf{k}}\\
12 & 1, \frac{1}{2}, {\bf k}, 1 & \tau_{\bf k}^+ c_{\bf k \uparrow}^\dagger  |0\rangle & 3\tilde\epsilon_{\bf{k}}+\frac{5U}{2}\\
13 & 1, -\frac{1}{2}, {\bf k}, 1 & \tau_{\bf k}^+ c_{\bf k \downarrow}^\dagger  |0\rangle & 3\tilde\epsilon_{\bf{k}}+\frac{5U}{2}\\
14 & 1, \frac{1}{2}, -{\bf k}, 1 & \tau_{\bf k}^+ c_{-{\bf k}\uparrow}^\dagger  |0\rangle & 3\tilde\epsilon_{\bf{k}}+\frac{5U}{2}\\
15 & 1, -\frac{1}{2}, -{\bf k}, 1 & \tau_{\bf k}^+ c_{-{\bf k}\downarrow }^\dagger |0\rangle & 3\tilde\epsilon_{\bf{k}}+\frac{5U}{2}\\
16 & 2, 0, 0, 0 & \frac{1}{2}\tau_{\bf k}^+ \tau_{\bf k}^+ |0\rangle & 4\tilde\epsilon_{\bf{k}}+5U\\ 
\hline
\end{array}
\]
\caption{Table of (normalized) basis states and associated $H_{\sf HK}$ energies. Here, $2M_{\bf k}+\nu_{\bf k}=N_{\bf k}$.}
\label{table5}
\end{table}
Based on the above, one can derive that the ground state of $H_{\sf HK}$ is of the form 
\begin{eqnarray}
|\Psi_0'\rangle = \prod_{{\bf k} \in \mathcal{S}_4} |2, 0,0,0\rangle_{\bf k}  \nonumber \otimes \prod_{{\bf k} \in \mathcal{S}_2'}  | \psi_{\bf{k}} \rangle  \otimes \prod_{{\bf k} \in \mathcal{S}_0}  | 0,0,0,0 \rangle_{\bf k} ,
\label{hkgroundstate}
\end{eqnarray}
with $\mathcal{S}_4$, $\mathcal{S}_0$, and $\mathcal{S}_2'=\mathcal{S}_3 \cup \mathcal{S}_2 \cup \mathcal{S}_1$ defined in Eq.\eqref{k-sectors}, and $|\psi_{\bf{k}} \rangle$ an element of the space spanned by the $N_{\bf k}=2$ states $|M_{\bf k},s^z_{\bf k},T_{\bf k},\nu_{\bf k}\rangle =|0, 1, 0, 2 \rangle$, $|0, -1, 0, 2\rangle$, $|0, 0, 0, 2\rangle$ and $|1, 0, 0, 0\rangle$ in Table \ref{table5}. Notice the conspicuous absence of the one-fermion and the three-fermion states in the ground state of the HK model, due to energetic considerations. This key feature of the HK model represents one of the main differences between our model and the HK model. This is illustrated for $d=1$ in Fig.  ~\ref{fig:nkhkd}. 

The resulting ground-state energy is given by 
\begin{equation}
E_{0}'(\mu, U) = \sum_{\bf k \in \mathcal{S}_4} \left( 4\tilde{\epsilon}_{\bf{k}}+ 5U  \right) +  \sum_{\bf k \in \mathcal{S}_2'} \left( 2\tilde{\epsilon}_{\bf{k}}  \right) .
\label{energyhk}
\end{equation}

From the expressions given above for the ground state and the ground-state energy, we see that the difference between the HK model and our model lies entirely in the two-particle sector. The HK model splits the two-particle sector into $4$ states of energy $2\tilde{\epsilon}_{\bf{k}}$ and $2$ states of energy $2\tilde{\epsilon}_{\bf{k}}+\frac{5}{2}U$, while our model splits the same set into $5$ states of energy $2\tilde{\epsilon}_{\bf{k}}+\frac{1}{2}U$ and one state of energy $2\tilde{\epsilon}_{\bf{k}}+\frac{5}{2}U$.

As demonstrated in Section \ref{Quantum Integrability}, our model is Richardson-Gaudin integrable in the presence of $s$-wave superconducting terms. The HK model is not. However, the HK model is separable at each BZ point. Our model is not. Our model is only separable at each pair of BZ points $(\bf{k},-\bf{k})$ taken together. Hence, we conclude that the price we pay for integrability in the presence of pairing terms is the entanglement of states at $\bf{k}$ and $-\bf{k}$. 

\subsection{Macroscopic Degeneracy}

The degeneracy of an arbitrary many-body eigenstate of the HK model is given by $4^{N_1+N_2+N_3}\,2^{N_2'}$, where $N_1 \, (N_3)$ is the number of momentum pairs $(\mathbf{k}, -\mathbf{k})$ present in the many-body state such that $ N_{\bf{k}} = 1 \, (3)$, $N_2$ is the number of momentum pairs $(\mathbf{k}, -\mathbf{k})$ such that $N_{\bf k}=2$ and $T_{\bf k}=0$ and $N_2'$ is the number of momentum pairs $(\mathbf{k}, -\mathbf{k})$ such that $N_{\bf k}=2$ and $T_{\bf k}\neq 0$. It is important to note that momentum pairs $(\mathbf{k}, -\mathbf{k})$ with $N_{\bf k}=0$ or 4 do not contribute to the macroscopic degeneracy of the many-body eigenstate. 

\subsection{One Spatial Dimension}

As we vary the value of the strength of the repulsion $U$ and the number of fermions (or equivalently, the chemical potential $\mu$) at zero temperature, we obtain a total of three metallic phases and one Mott-insulating phase. These are depicted in Fig. ~\ref{fig:phases} of the main text.

Each of the metallic phases is labeled based on the nature of the occupancies of the energy levels in the ground state, as in our model. However, since the allowed two-particle states in the HK ground states differ from those in the ground states of  $H_{\sf n}$, we use \textquotedblleft(2')\textquotedblright for clarity. The free Fermi-liquid is located at precisely $U=0$.
The Mott-insulating phase is observed along a line at half-filling beyond a critical value of $U$. This phase is labeled \textquotedblleft(2')\textquotedblright.  This phase is seen in the $\rho_F$ vs $U$ phase diagram as a first order phase transition between the phases \textquotedblleft(2',0)\textquotedblright and \textquotedblleft(4,2')\textquotedblright, see Fig. \ref{fig:phases}. 

Given the above, it is straightforward to calculate the location of the quantum phase transitions as a function of $U$ and $\mu$ just as in our model. The phase transition in $d=1$ between \textquotedblleft(4,2',0)\textquotedblright and \textquotedblleft(2',0)'' is at $\mu = \frac{5U}{2} -1$ and that between \textquotedblleft(4,2',0)\textquotedblright and \textquotedblleft(4,2')\textquotedblright is at $\mu=1$. Additionally, the Mott-insulating phase ``(2')'' is found at $\rho_F=1$, and $U> \frac{4}{5}$.

\subsection{Higher Spatial Dimensions}

In $d$ spatial dimensions, via similar arguments, we see the same three metallic phases and one Mott-insulating phase \textquotedblleft(2')\textquotedblright. The location of the quantum phase transitions between \textquotedblleft(4,2',0)\textquotedblright and \textquotedblleft(2',0)'' is at $ \mu = \frac{5U}{2}-d$,  and that between \textquotedblleft(4,2',0)\textquotedblright and \textquotedblleft(4,2')'' is at $\mu = d$.  The Mott-insulating phase \textquotedblleft(2')\textquotedblright (which appears as a first-order phase transition in the $\rho_F$ vs $U$ phase diagram) is located at half-filling ($\rho_F=1$) for $U> \frac{4d}{5}$ (see Fig. \ref{fig:phasesmu-rho}). 

Many-body Lifshitz transitions are also seen criss-crossing the phase diagram for the same reasons as in our model. If a Lifshitz transition occurs in the free Fermi liquid at $\mu= \mu_0$ (where $\mu_0$ is a constant), then this transition splits into two many-body Lifshitz transitions and extends into the phase diagram along the lines $\mu-\mu_0= \frac{5U}{2}$ and  $\mu-\mu_0= 0$.

\subsection{Charge Gap}

It can be proved that the HK model has a vanishing charge gap, and hence describes a metal, at any value of $\rho_F$ except at half filling, i.e., $\rho_F=1$. At $\rho_F =  1$, we obtain
\begin{eqnarray}
   \Delta \mu (\rho_F)  =
\begin{cases}
\text{0}, & \text{if $ U<\frac{4d}{5}$}, \\
\text{$\frac{5U}{2}-2d $}, & \text{if $U>\frac{4d}{5}$},
\end{cases}
\end{eqnarray}
thus motivating the identification of the phase \textquotedblleft(2')\textquotedblright as a Mott-insulating phase.

\subsection{Retarded Green's Function}

For the HK model,  the retarded Green's function is given by
\begin{equation}
G_{ \sigma}({\bf{k}}, \omega) = \frac{1-\left< n_{\bf{k} \bar{\sigma}}\right>}{\omega -\tilde{\epsilon}_{\bf{k}}} + \frac{\left< n_{\bf{k} \bar{\sigma}}\right>}{\omega - \tilde{\epsilon}_{\bf{k}}-\frac{5}{2}U} .
\end{equation}

It should be noted that both poles of the HK model are also poles of our model. However, our model has two additional poles. On the other hand, just as in our model, location of the poles in the half-BZ coincide with the location of the many-body Fermi surfaces. 

\section{Green’s Function of our Model}
\label{AppendixB}

We want to calculate $\left< \left[c^{\;}_{\bf{k} \sigma}(t), c^{\dagger}_{\bf{k} \sigma}(0)\right]_+   \right>$, at time $t>0$, for our model Hamiltonian $H_{\sf n}$. 
It can be proved that 
\begin{eqnarray}
\left[ H_{\sf n}, c_{\bf{k} \sigma} \right] &=& \tilde{M}_{11} c_{\bf{k} \sigma} + \tilde{M}_{1 2} d_{\bf{k} \sigma} \nonumber ,\\
\left[ H_{\sf n}, d_{\bf{k} \sigma} \right] &=& \tilde{M}_{21} c_{{\bf{k}} \sigma} + \tilde{M}_{2 2} d_{\bf{k} \sigma},
\end{eqnarray}
where $d_{\bf{k} \sigma} = (\delta_{\sigma \downarrow}-\delta_{\sigma \uparrow})\tau_{\bf{k}}^- c_{-{\bf{k}} \bar{\sigma}}^{\dagger}$ and $\tilde{M}_{ij}$ are elements of a $2 \times 2$ matrix $\tilde{M}$ of operators 
\begin{eqnarray}
\tilde{M}_{11} &=& - \tilde{\epsilon}_{\bf{k}} -U(\tau_{\bf{k}}^z+2) \nonumber ,\\
\tilde{M}_{12} &=& U \nonumber ,\\
\tilde{M}_{21} &=& U \tau_{\bf{k}}^- \tau_{\bf{k}}^+ \nonumber ,\\
\tilde{M}_{22} &=& - \tilde{\epsilon}_{\bf{k}} +U(\tau_{\bf{k}}^z-1) .
\end{eqnarray}

Notice that each of the elements of $\tilde{M}$ commute with the Hamiltonian. 
Next, let us introduce a column vector ${\sf v}$ whose top entry is $c_{\bf{k} \sigma}$ and bottom entry is $d_{\bf k \sigma}$. Since $H_{\sf n}$ commutes with each $\tilde{M}_{ij}$, we can see that $(\mathcal{H}_{\sf n})^n {\sf v}= \tilde{M}^n {\sf v}$, where $(\mathcal{H}_{\sf n})^n {\sf v} = [H_{\sf n},[H_{\sf n},[H_{\sf n},\cdots, [H_{\sf n},{\sf v}]\cdots]]]$. 
In order to simplify calculations, it is useful to write $\tilde{M}$ as $\tilde{M}= U M -\lambda \mathcal{I}$, where $\lambda = \tilde{\epsilon}_{\bf{k}} + \frac{5}{4}U$, and $\mathcal{I}$ is the identity operator.  Consequently, the elements of $M$ are given by
\begin{eqnarray}
M_{11} &=& -\tau^z_{\bf{k}} -\frac{3}{4} \nonumber ,\\
M_{12} &=&1 \nonumber ,\\
M_{21} &=& \tau_{\bf{k}}^- \tau_{\bf{k}}^+ \nonumber ,\\
M_{22} &=& \tau_{\bf{k}}^z+ \frac{1}{4}.
\end{eqnarray}
We intend to use the equations of motion method to calculate the Green's function. In this method, we first deal with the equal time commutators of $c_{\bf{k} \sigma}$ with the Hamiltonian in order to calculate $c_{\bf k \sigma}(t)$.

Given an arbitrary operator (in the Heisenberg representation) $A(t)$, denote the Fourier transform  w.r.t. time of the quantity $-i \theta(t)\left< [A(t), c^{\dagger}_{\bf{k} \sigma}(0)]_+   \right>$ as $\widehat{A}(\omega)$, and the expectation value of the equal-time anticommutator $\left< \left[A(0), c^{\dagger}_{\bf{k} \sigma}(0)  \right]_+ \right>=\bar A$. Therefore, the required Green's function in momentum and frequency space, $G_\sigma({\bf k},\omega)$, is the top entry of $\widehat{\sf v}(\omega)$.

It can be proved that for any operator $A$, $\omega \widehat{A}(\omega) = \bar{A} -\widehat{B}(\omega)$, where  $ B(t)= \left[H_{\sf n}, A(t)  \right]$. It can be seen that this identity gives us one relation for $\widehat{{\sf v}}(\omega)$ in terms of $\bar{\sf v}$ and $\widehat{\tilde{M}{\sf v}}\left(\omega\right)$ and another relation for $\widehat{\tilde{M}{\sf v}}\left( \omega  \right)$ in terms of $\overline{ \tilde{M}{\sf v}}$ and $\widehat{\tilde{M}^2{\sf v}}\, (\omega)$.

We can then express $\tilde{M}$ in terms of $M$ and simplify. In order to simplify the resulting term $\widehat{M^2{\sf v}}\, (\omega)$, it can be proved that $M$ satisfies the equation $M^2 + \frac{1}{2}M -\left({\cal C}^2_{\bf k} + \frac{3}{16}\mathcal{I}\right)= 0$, where ${\cal C}^2_{\bf k}$ is the Casimir of the pseudo-spin algebra. We can use this relation to write $M^2$ in terms of $M$ and ${\cal C}^2_{\bf k}$ and simplify to get
\begin{eqnarray}
\left(\omega^2 -\omega(\frac{1}{2}U+ 2\lambda) -\frac{3}{16}U^2 + \lambda^2 + \frac{1}{2}\lambda U  \right)\widehat{{\sf v}}(\omega) \nonumber \\
\!\!= (\omega -\frac{1}{2}U -\lambda)\bar{ {\sf v}} -U \overline{M{\sf v}} + U^2\, \widehat{{\cal C}^2_{\bf k}{\sf v}}\,(\omega) .
\end{eqnarray}
We emphasize that the above equation reduces the calculation of $\widehat{{\sf v}}(\omega)$ to that of $\widehat{{\cal C}^2_{\bf k}{\sf v}}\,(\omega)$. 

We can repeat the equation of motion procedure used to determine $\widehat{{\sf v}}(\omega)$, but now applied to $\widehat{{\cal C}^2_{\bf k}{\sf v}}\, (\omega)$ and $(\widehat{{\cal C}^2_{\bf k})^2{\sf v}}\, (\omega)$ yielding 
\begin{eqnarray}
\hspace*{-2.cm}&&
\left(\omega^2 -\omega(\frac{1}{2}U+ 2\lambda) -\frac{3}{16}U^2 + \lambda^2 + \frac{1}{2}\lambda U  \right)\widehat{{\cal C}^2_{\bf k}{\sf v}}\,(\omega) \nonumber \\
\hspace*{-0.7cm}&&= (\omega -\frac{1}{2}U -\lambda)\overline{ {\cal C}^2_{\bf k}{\sf v}} -U \, \overline{{\cal C}^2_{\bf k}M{\sf v}} + U^2\, \widehat{({\cal C}^2_{\bf k})^2{\sf v}}\,(\omega), \\
\hspace*{-2.cm}&&\left(\omega^2 -\omega(\frac{1}{2}U+ 2\lambda) -\frac{3}{16}U^2 + \lambda^2 + \frac{1}{2}\lambda U  \right)\widehat{({\cal C}^2_{\bf k})^2{\sf v}}\, (\omega) \nonumber \\ \hspace*{-0.7cm}&&=(\omega -\frac{1}{2}U -\lambda)\overline{ ({\cal C}^2_{\bf k})^2{\sf v}} -U \, \overline{({\cal C}^2_{\bf k})^2M{\sf v}} + U^2\, \widehat{({\cal C}^2_{\bf k})^3{\sf v}}\,(\omega) .
\label{recursion}
\end{eqnarray}

Let us now turn to the operator ${\cal C}^2_{\bf k}$. This takes the value $2$, $\frac{3}{4}$ and $0$ when acting on the states with $\nu_{\bf k}=0$,$1$ and $2$,  respectively. Next, consider the operator  ${\cal C}^2_{\bf k}({\cal C}^2_{\bf k}-\frac{3}{4}\mathcal{I})({\cal C}^2_{\bf k}-2\mathcal{I})$. Notice that it vanishes when acting on any state in the Hilbert space. Hence, we can write
\begin{equation}
{\cal C}^2_{\bf k}\left({\cal C}^2_{\bf k}-\frac{3}{4}\mathcal{I}\right)\left({\cal C}^2_{\bf k}-2\mathcal{I}\right) =0 .
\label{closure}
\end{equation}
This allows us to express $({\cal C}^2_{\bf k})^3$ in terms of lower powers of ${\cal C}^2_{\bf k}$. This immediately tells us that the series generated by the equations of motion method closes at sixth order! Hence, using Eqs. \eqref{closure} and \eqref{recursion}, we can write
\begin{eqnarray}
&&\widehat{({\cal C}^2_{\bf k})^2{\sf v}}\,(\omega) =  \\ &&\frac{(\omega -\frac{1}{2}U -\lambda)\overline{ ({\cal C}^2_{\bf k})^2{\sf v}} -U \overline{({\cal C}^2_{\bf k})^2M{\sf v}}  -\frac{3}{2}U^2\, \widehat{{\cal C}^2_{\bf k}{\sf v}}\, (\omega)}{\omega^2 -\omega(\frac{1}{2}U+ 2\lambda) -\frac{47}{16}U^2 + \lambda^2 + \frac{1}{2}\lambda U} \nonumber
\end{eqnarray}
We can subsequently use this expression to obtain a closed-form expression for $\widehat{{\sf v}}(\omega)$. The final result is 
\begin{eqnarray}
\widehat{{\sf v}}(\omega) &=& \frac{A_1}{\omega- \tilde{\epsilon}_{\bf{k}}}+ \frac{A_2}{\omega- \tilde{\epsilon}_{\bf{k}}-\frac{1}{2}U} + \frac{A_3}{\omega- \tilde{\epsilon}_{\bf k}-2U} \\ &&+ \frac{A_4}{\omega- \tilde{\epsilon}_{\bf k}-\frac{5}{2}U} + \frac{A_5}{\omega- \tilde{\epsilon}_{\bf k}-U} + \frac{A_6}{\omega- \tilde{\epsilon}_{\bf k}-3U} \nonumber, 
\end{eqnarray}
where the two-component vectors are
\begin{eqnarray}
A_1 &=& \frac{1}{120}\bigg[-21 \,\overline{{\cal C}^2_{\bf k}{\sf v}} - 12 \,\overline{{\cal C}^2_{\bf k}M{\sf v}} + 28 \,\overline{  ({\cal C}^2_{\bf k})^2{\sf v}}  \nonumber\\&&+ 16 \, \overline{ ({\cal C}^2_{\bf k})^2M{\sf v}} \bigg], \nonumber \\
A_2 &=& \frac{2}{15} \bigg[10 \, \overline{{\cal C}^2_{\bf k}{\sf v}} + 8 \,\overline{{\cal C}^2_{\bf k}M{\sf v}} - 5 \,\overline{ ({\cal C}^2_{\bf k})^2{\sf v}} \nonumber \\ &&- 4 \, \overline{ ({\cal C}^2_{\bf k})^2M{\sf v}} \bigg], \nonumber \\
A_3 &=& \frac{1}{24} \bigg[6 \, \overline{ {\sf v}} - 24 \, \overline{ M{\sf v}} - 11 \,\overline{{\cal C}^2_{\bf k}{\sf v}} + 44 \,\overline{{\cal C}^2_{\bf k}M{\sf v}} \nonumber \\ &&+ 4 \, \overline{  ({\cal C}^2_{\bf k})^2{\sf v}} - 16 \, \overline{  ({\cal C}^2_{\bf k})^2M{\sf v}} \bigg], \nonumber \\ 
A_4 &=& \frac{2}{15} \bigg[  6 \, \overline{{\cal C}^2_{\bf k}{\sf v}} - 8 \,\overline{{\cal C}^2_{\bf k}M{\sf v}} - 3 \, \overline{ ({\cal C}^2_{\bf k})^2{\sf v}} \nonumber \\ &&+ 4 \, \overline{ ({\cal C}^2_{\bf k})^2M{\sf v}} \bigg] , \nonumber \\
A_5 &=& \frac{1}{24} \bigg[ 18 \, \overline{\sf v} + 24 \, \overline{M{\sf v}} - 33 \, \overline{{\cal C}^2_{\bf k}{\sf v}}- 44 \, \overline{{\cal C}^2_{\bf k}M{\sf v}}\nonumber \\ &&+ 12 \,\overline{ ({\cal C}^2_{\bf k})^2{\sf v}} + 16 \,\overline{ ({\cal C}^2_{\bf k})^2M{\sf v}} \bigg] , \nonumber \\
A_6 &=& \frac{1}{120}\bigg[  -15 \,\overline{{\cal C}^2_{\bf k}{\sf v}} + 12 \,\overline{{\cal C}^2_{\bf k}M{\sf v}} + 20 \,\overline{ ({\cal C}^2_{\bf k})^2{\sf v}} \nonumber \\ &&- 16 \, \overline{({\cal C}^2_{\bf k})^2M{\sf v}}  \bigg] . 
\end{eqnarray}

After a straightforward evaluation of the quantities above, one can show that the residues $A_5$ and $A_6$ vanish for the top entry of ${\sf v}$, leaving us with the four poles which are explicitly written down in the main paper. 

\section{Cooper Pair Instability}
\label{Onepairapp}

We aim to study the behavior of the roots $x$ of the Richardson equation in $d=1$ for a single pairon, and subsequently derive a rigorous bound for the smallest root $x=\tilde x$ of the Cooper pair equation, including repulsion
\begin{equation}
\sum_{k \in \mathcal{S}_0}\frac{1}{\tilde{\epsilon
}_{k}-x} + \frac{1}{2}\sum_{k \in \mathcal{S}_1}\frac{1}{\tilde{\epsilon
}_{k}-x}=\frac{1}{G},
\end{equation}
where the sets ${\cal S}_0$ and ${\cal S}_1$ are defined in Eq. \eqref{k-sectors}.



First, we observe that the roots of the above equation are always real for any finite positive value of $G$. Next, in order to make progress, let us define the function $f(y)$ as the LHS of the above equation, i.e.,
\begin{equation}
f(y)=\sum_{k \in \mathcal{S}_0}\frac{1}{\tilde{\epsilon
}_{k}-y} + \frac{1}{2}\sum_{k \in \mathcal{S}_1}\frac{1}{\tilde{\epsilon
}_{k}-y} .
\end{equation}

The following properties of $f(y)$ can be readily verified:
\begin{itemize}
    \item $f(y)$ is continuous and differentiable at all $y$ except at the set of isolated singular points $y=\tilde{\epsilon}_k$ where $k\in\mathcal{S}_0 \cup \mathcal{S}_1$.
    \item The behavior of $f(y)$ near the singular points is given by  $\lim_{y \rightarrow \tilde{\epsilon}_k \pm 0^+} f(y) = \mp \infty$.
    \item $f(y)$ is {\it monotonically} increasing in the interval $(\tilde \epsilon_{k_a}, \tilde \epsilon_{k_b})$  between two {\it neighboring} singular points.
    \item $f(y)$ is {\it monotonically} increasing in the intervals $(-\infty, \text{min}_{k}\tilde{\epsilon}_{k} )$ and $(\text{max}_k \tilde{\epsilon}_k, \infty )$, where minimization and maximization are over the set $ \mathcal{S}_0 \cup \mathcal{S}_1 $.
\end{itemize}

Consequently,  for any finite positive value of $G$:
\begin{itemize}
    \item The total number of roots of the Richardson equation is equal to $|\mathcal{S}_0|+|\mathcal{S}_1|$. Hence, the roots $x=x^{(k)}$ are in a one-to-one correspondence with the elements of the set $\mathcal{S}_0 \cup \mathcal{S}_1$.
    \item There exists exactly one root between two {\it neighboring} singular points of $f(y)$. 
    \item The smallest root, $x=\tilde{x}$, lies in the interval $(-\infty, \text{min}_{k}\tilde{\epsilon}_{k} )$.
\end{itemize}

To determine which root should be selected, we now consider the minimization of \( E(\mu, U, G, M = 1) \). It follows that, among all the roots \( x^{(k)} \), the appropriate choice is the smallest one, \( x = \tilde{x} \). It is straightforward to verify that $\text{min}_{k}\tilde{\epsilon}_{k} =\tilde{\epsilon}_2$ and $\text{max}_{k}\tilde{\epsilon}_{k}=1-\mu$. Consequently, by using the monotonic behavior of $f(y)$ in the interval $(-\infty, \tilde{\epsilon}_2)$ and the fact that $\lim_{y\rightarrow\pm\infty} f(y) =0$, one can show that, as $G$ is tuned from $+\infty$ down to $0$ with $\mu$ and $U$ held fixed, $\tilde{x}$ increases continuously and {\it monotonically} from $-\infty$ up to $\tilde{\epsilon}_2$. 

To derive more stringent bounds for $\tilde{x}$, let us consider the trivial inequality $\tilde{\epsilon}_2 \leq \tilde{\epsilon}_k\leq 1-\mu$, which is valid for all $k \in \mathcal{S}_0 \cup \mathcal{S}_1$, and focus on the interval $y\in (-\infty, \tilde{\epsilon}_2)$. This allows us to write 
\begin{equation}
\frac{1}{\tilde{\epsilon}_2-y} \geq \frac{1}{\tilde{\epsilon}_k-y} \geq \frac{1}{1-\mu-y}. 
\end{equation}
Summing over  the sets $\mathcal{S}_0$ and $\mathcal{S}_1$ separately, we obtain
\begin{equation}
\frac{|\mathcal{S}_i|}{\tilde{\epsilon}_2-y}\geq \sum_{k \in \mathcal{S}_i}\frac{1}{\tilde{\epsilon}_k-y} \geq \frac{|\mathcal{S}_i|}{1-\mu-y},
\end{equation}
for $i=0,1$, which leads us to 
\begin{equation}
\frac{Q_{\sf aux}}{\tilde{\epsilon}_2-y}-\frac{1}{G}\geq f(y) -\frac{1}{G} \geq \frac{Q_{\sf aux}}{1-\mu-y} - \frac{1}{G},
\end{equation}
where $Q_{\sf aux}= |\mathcal{S}_0|+\frac{1}{2}|\mathcal{S}_1|=\frac{1}{2}(V-N)+1$ and $N$ is the total number of fermions in the system. Next, notice that the three expressions in the inequality above are {\it monotonically} increasing continuous functions of $y$. Evaluating the above inequality at $y=\tilde{x}$, we obtain the required bound for $\tilde{x}$
\begin{equation}
\tilde{\epsilon}_2 - Q_{\sf aux}G \leq \tilde{x} \leq 1-\mu- Q_{\sf aux}G.
\end{equation}

In higher dimensions, the one-to-one correspondence between the elements of $\mathcal{S}_0 \cup \mathcal{S}_1$ and the roots of the Richardson equation no longer holds due to possible degeneracies in $\tilde{\epsilon}_{\mathbf{k}}$. Nevertheless, a straightforward generalization of the above proof yields the following bounds for $\tilde{x}$
\begin{equation}
\tilde{\epsilon}_2 - Q_{\sf aux}G \leq \tilde{x} \leq d-\mu- Q_{\sf aux}G.
\end{equation}

Finally, to determine the asymptotic behavior of $\tilde{x}$ in the limit $GV \gg 1$, we assume that all ``charges'' located at $\tilde{\epsilon}_{\mathbf{k}}$ with $\mathbf{k} \in \mathcal{S}_0 \cup \mathcal{S}_1$ are effectively concentrated at their ``center of charge". Under this approximation we obtain 
\begin{equation}
\tilde{x} (GV \gg 1) \sim \tilde{x}_c - Q_{\sf aux}G, 
\end{equation}
where the quantity $\tilde{x}_c$ (the ``center of charge") depends on $\mu$ and $U$, but is independent of $G$, and is given by
\begin{equation}
\tilde{x}_c = \frac{1}{Q_{\sf aux}} \left( \sum_{\bf k \in \mathcal{S}_0} \tilde{\epsilon}_{\bf k}+ \frac{1}{2}\sum_{\bf k \in \mathcal{S}_1} \tilde{\epsilon}_{\bf k}  \right) .
\end{equation}

\bibliography{references}

\end{document}